%% file: main_arxiv.tex
\documentclass[10pt,letterpaper,compsoc,conference]{iiswc23}

\usepackage{cite}
\usepackage{amsmath,amssymb,amsfonts}
\usepackage{algorithmic}
\usepackage{graphicx}
\usepackage[dvipsnames]{xcolor}
\usepackage[final]{microtype}
\usepackage[italic]{mathastext}
\usepackage{libertine}
\usepackage[T1]{fontenc}
\usepackage{textcomp}
\usepackage[varqu,varl]{zi4}
\usepackage[all]{nowidow}
\usepackage[auth-lg,affil-it]{authblk}
\usepackage[keeplastbox]{flushend}

\usepackage{paralist}

\usepackage{booktabs}

\usepackage{booktabs}
\usepackage{import}
\usepackage{multirow}
\usepackage{pgfplots}
\pgfplotsset{compat=1.16}
\usepackage{subcaption}
\usepackage{xspace}
\usepackage[vlined,ruled]{algorithm2e}
\usepackage{url}
\usepackage{tikz}
\usetikzlibrary{matrix, plotmarks, chains, positioning, shapes.multipart, calc, shapes, arrows, shapes.arrows}
\usetikzlibrary{calc,decorations.pathreplacing,calligraphy}
\usepackage{pgf}
\usepackage{pgfplots}
\pgfplotsset{compat=1.16}
\usepackage[binary-units=true]{siunitx}

\newcommand{\bhivel}{\emph{BHive\textsubscript{L}}\xspace}
\newcommand{\bhiveu}{\emph{BHive\textsubscript{U}}\xspace}
\newcommand{\uiCA}{uiCA\xspace}
\newcommand{\microops}{{\textmu}ops\xspace}
\newcommand{\microop}{{\textmu}op\xspace}
\newcommand{\negvspace}{\vspace{-1.5mm}}
\renewcommand{\negvspace}{}

\usepackage{pdfcomment}
\newcommand{\todo}[1]{\marginpar{\pdfcomment[color=red]{Todo: #1}}}
\renewcommand{\todo}[1]{}

\newcommand{\DefTPL}{\ensuremath{TP_L}\xspace}
\newcommand{\DefTPU}{\ensuremath{TP_U}\xspace}

\newcommand{\wrongDef}[1]{\textcolor{gray}{#1}}

\newcommand\copyrighttext{%
\footnotesize \textcopyright\ 2023 IEEE. Personal use of this material is permitted. Permission from IEEE must be obtained for all other uses, in any current or future media, including reprinting/republishing this material for advertising or promotional purposes, creating new collective works, for resale or redistribution to servers or lists, or reuse of any copyrighted component of this work in other works.
}
\def\mycopyrightnotice{
\begin{tikzpicture}[remember picture,overlay]
\node[anchor=south,yshift=10pt] at (current page.south) {\fbox{\parbox{\dimexpr\textwidth-\fboxsep-\fboxrule\relax}{\copyrighttext}}};
\end{tikzpicture}%
}
\makeatletter
\def\ps@IEEEtitlepagestyle{
  \def\@oddfoot{\mycopyrightnotice}
  \def\@evenfoot{}
}

\begin{document}

\newcommand{\toolname}{\textsc{Facile}\xspace}


\title{\toolname: Fast, Accurate, and Interpretable\\Basic-Block Throughput Prediction}


\renewcommand\Authsep{\qquad}
\renewcommand\Authand{\qquad}
\renewcommand\Authands{\qquad}


\author{Andreas Abel}
\author{Shrey Sharma}
\author{Jan Reineke}
\affil{Saarland University\\ Saarland Informatics Campus \\ Saarbr\"ucken, Germany}


\maketitle


\begin{abstract}
Basic-block throughput models such as uiCA, IACA, GRANITE, Ithemal, llvm-mca, OSACA, or CQA guide optimizing compilers and help performance engineers identify and eliminate bottlenecks.
For this purpose, basic-block throughput models should ideally be fast, accurate, and interpretable.

Recent advances have significantly improved accuracy:
uiCA, the state-of-the-art model, achieves an error of about 1\% relative to measurements across a wide range of microarchitectures.
The computational efficiency of throughput models, which is equally important for widespread adoption, especially in compilers, has so far received little attention.

In this paper, we introduce \toolname, an analytical throughput model that is fast, accurate, and interpretable.
\toolname analyzes different potential bottlenecks independently and analytically.
Due to its compositional nature, \toolname's predictions directly pinpoint the bottlenecks.
We evaluate \toolname on a wide range of microarchitectures and show that it is almost two orders of magnitude faster than existing models while achieving state-of-the-art accuracy.
\end{abstract}

\section{Introduction}

Basic-block throughput models predict the steady-state throughput of basic blocks on a given microarchitecture.
Such models are used by compilers~\cite{Schkufza2013,Schkufza2016,Mangpo2016,Souper2017,Minotaur2023,Trofin21,Stephenson03,McGovern98} to guide optimizations and by performance engineers to pinpoint bottlenecks, which can subsequently be mitigated by code changes or improvements to the microarchitecture.

Ideally, basic-block throughput models should be accurate, efficient, and interpretable:
\begin{compactitem}
\item The utility of throughput predictions clearly depends on their accuracy.
The inaccuracy of models may misguide optimizations and lead to suboptimal performance~\cite{Mendis18,Pohl19,Pohl20}.\looseness=-1

\item Depending on the use case, the speed of the model is equally important.
For example, superoptimizers~\cite{Schkufza2013,Schkufza2016,Mangpo2016,Souper2017,Minotaur2023} explore a vast space of possible instruction sequences to find the fastest implementation of a given functionality.
In such scenarios, the speed of the model is a limiting factor for the overall performance of the superoptimizer.

\item Finally, interpretability is important to understand the underlying reasons for the performance of a basic block, which can help performance engineers or compilers to pick appropriate optimizations.
\end{compactitem}

Recently, several basic-block throughput models have been introduced including uiCA~\cite{Abel22}, IACA~\cite{iacaGuide}, llvm-mca~\cite{dibiagio18,llvmmca}, OSACA~\cite{Laukemann18,Laukemann19}, CQA~\cite{cqa14}, GRANITE~\cite{sykora22}, Ithemal~\cite{mendis19a}, and DiffTune~\cite{Renda20}.
These models offer different trade-offs between accuracy, speed, and interpretability.
When it comes to accuracy, the state-of-the-art is defined by uiCA~\cite{Abel22}, a simulation-based model:
On the BHive benchmark suite~\cite{Chen19} it achieves an average error of around 1\%, while the average error of the other models is about an order of magnitude higher.
On the other hand, being simulation-based, uiCA is among the slower models.
It is also not straightforward to determine the reasons behind the performance of a basic block from uiCA's cycle-accurate simulation outputs.
The fastest available model Ithemal~\cite{mendis19a}, a machine-learned (ML) model, takes about 10~ms on the average to predict the throughput of a basic block, which is an order of magnitude faster than uiCA.
However, due to its ML-nature Ithemal is not easily interpretable.
\todo{discuss interpretability of existing models}

In this paper, we introduce \toolname, a basic-block throughput model that is fast, accurate, and interpretable.

\toolname is based on the following hypothesis: the throughput of a basic block is determined by the maximum of a small set of potential bottlenecks that can be analyzed independently of each other.
These potential bottlenecks correspond to pipeline components in the front end and the back end of the microarchitecture, as well as to precedence constraints among instructions in successive loop iterations.
We introduce efficient analytical predictors for each of these components, which are combined to predict the throughput of a basic block and to identify the bottlenecks that limit its performance in Section~\ref{sec:predictor}.\looseness=-1

After briefly discussing \toolname's implementation in Section~\ref{sec:implementation}, we extensively evaluate \toolname's accuracy, speed, and interpretability in Section~\ref{sec:evaluation}.
It achieves an average error comparable to uiCA, while being nearly two orders of magnitude faster than Ithemal.
It is also more interpretable than both uiCA and Ithemal, as it is based on a simple analytical model.
We exploit \toolname's interpretability to gain insights into the evolution of Intel microarchitectures and the impact of individual components on performance.\looseness=-1

To summarize, we make the following contributions:
\begin{compactitem}
\item We\hspace{-0.2mm} introduce\hspace{-0.2mm} \toolname,\hspace{-0.2mm} an analytical\hspace{-0.25mm} basic-block throughput model that is fast, accurate, and interpretable.
\item We extensively evaluate \toolname's accuracy and speed demonstrating that it achieves state-of-the-art accuracy while being almost two orders of magnitude faster than the fastest available models to date. 
\item As a minor contribution, we exploit \toolname's interpretability to gain insights into the evolution of Intel microarchitectures.
\end{compactitem}

\toolname is available open source as part of the uiCA repository at \url{https://github.com/andreas-abel/uiCA}.
The Python file \url{facile.py} acts as the front end to the tool.

\todo{can we say something about why \toolname is so much faster than the other models? what are they doing wrong?}


\todo{have simple table or graph that shows how accurate and fast our model is; possibly also something about interpretability}

\section{Related Work}

Existing throughput predictors can be broadly classified into three categories: simulation-based, analytical, and learning-based models.

Simulation-based throughput predictors model the complete CPU microarchitecture or parts of it to predict the throughput of a basic block based on cycle-by-cycle simulation.
Examples include tools like the uops.info Code Analyzer~(uiCA)~\cite{Abel22}, the LLVM Machine Code Analyzer~(llvm-mca)~\cite{dibiagio18,llvmmca}, and the Code Quality Analyzer~(CQA)~\cite{cqa14}.
llvm-mca uses the LLVM compiler framework's~\cite{llvm04} scheduling models to predict the performance of machine code.
It does not model the front end of a processor pipeline or techniques like macro or micro fusion, which can affect the performance of a basic block.
On the other hand, CQA uses a detailed model of the front end of the CPU pipeline, but does not model the back end of the pipeline ``because of its complexity and lack of documentation''~\cite{cqa14}.
Finally, uiCA, the current state-of-the-art in terms of accuracy, models both the front end and the back end of modern Intel CPU pipelines, as well as techniques like macro and micro fusion or move elimination at a high level of detail.
\toolname is similar to uiCA in that it models both the front end and the back end of the CPU pipeline, as well as techniques like macro and micro fusion but it does not perform a cycle-by-cycle simulation of the basic block's execution, rather it considers different pipeline components independently.

To reduce the manual effort that is typically required to set the microarchitecture-specific parameters of simulation-based throughput predictors, DiffTune~\cite{Renda20} uses a neural network-based technique to learn the parameters of the llvm-mca simulator automatically.
\cite{Abel22b} proposes a significantly simpler learning algorithm for the same setting that can serve as a baseline, and showed that DiffTune does not beat this baseline.

Unlike simulation-based basic block throughput predictors, full-system simulators~\cite{austin02,Binkert11,Loh09,patel11,sanchez13,Carlson11,yourst07} model the complete system with all its hardware components and the intricate interactions between them.
Such simulators are used to evaluate the performance of entire programs rather than individual basic blocks.

Analytical throughput predictors, such as the Open Source Architecture Code Analyzer (OSACA)~\cite{Laukemann18,Laukemann19} and \toolname, in contrast, do not rely on cycle-by-cycle simulation, but use analytical formulas and more coarse-grained models of the microarchitecture for throughput prediction.
We note though that the term ``analytical'' is broad and analytical models can vary significantly in their level of detail and may even include some form of simulation.

Learning-based throughput predictors use machine learning techniques to predict the throughput of a basic block.
They try to alleviate the need for detailed and tedious manual modeling of the microarchitecture and its interactions with the program by learning performance models directly from measured data.
Also, learning-based models can in principle be easily adapted to new microarchitectures.
Ithemal~\cite{mendis19a} is one such predictor that uses an LSTM-based neural network to predict the throughput of a basic block.
More recently a graph neural network based predictor called GRANITE~\cite{sykora22} was proposed.
A significant drawback of learning-based predictors is that they are not interpretable.
They are often referred to as ``black boxes'' because it is difficult to understand how they arrive at their predictions.
In contrast, \toolname is based on a simple analytical model that is easy to understand and interpret.\looseness=-1

To understand the predictions of learning-based predictors, the CoMEt~\cite{chaudhary23} framework proposes using post-hoc explanation techniques like the Anchor's algorithm~\cite{ribeiro18}.
The goal being to explain the predictions of a black-box model by identifying instructions in the input basic block that are most relevant to the prediction.
Similarly, AnICA~\cite{Ritter22} uses differential testing and abstract interpretation to generate ``abstract basic blocks'' that explain inconsistencies between throughput predictors.
\toolname primarily provides insights into the impact of individual components of the pipeline on the basic-block performance as opposed to identifying instructions that influence its performance.



\section{Background}

\subsection{Basic-Block Throughput Notions}

The reciprocal throughput (also known as the ``gap'' or the ``inverse throughput'' in the literature) of a basic block is defined as the number of clock cycles required per iteration of the basic block when it is executed repeatedly in a steady state.
As is common in the literature on basic-block throughput prediction, we refer to the reciprocal throughput simply as the ``throughput'' in the remainder of this paper.

However, this basic definition can be interpreted in different ways depending on the type of basic block that is being considered.
Consider a basic block that ends in a branch instruction to the beginning of the block.
In this case, ``executing the basic block repeatedly'' can reasonably be interpreted as executing the basic block repeatedly in an infinite loop.
In the following, we will refer to the corresponding notion of throughput as \DefTPL.
On the other hand, consider a basic block that does not end in a branch instruction.
In this case, ``executing the basic block repeatedly'' can reasonably be interpreted as unrolling the basic block for a significant number of iterations.
In the following, we will refer to the throughput in this case as \DefTPU.\looseness=-1

It is important to distinguish between \DefTPL and \DefTPU,
as the pipeline components used by basic blocks are different in these two cases.
For a basic block that ends in a branch instruction, in steady state, the decoded \microops are typically streamed from the \textmu OP cache (also called Decoded Stream Buffer, short DSB) or the loop stream detector (LSD).
Unrolled basic blocks, on the other hand, are typically fetched from the instruction cache (IC) and decoded by the decoders.\looseness=-1

Previous work did not always clearly distinguish these two notions of throughput~\cite{Abel22}.
Intel's IACA~\cite{IACA} is based on \DefTPL, and so are OSACA~\cite{Laukemann18, Laukemann19} and CQA~\cite{cqa14}.
On the other hand, Ithemal~\cite{mendis19a} and DiffTune~\cite{Renda20} are based on \DefTPU.
For llvm-mca~\cite{llvmmca}, it is not so clear. As llvm-mca does not model potential bottlenecks in the front end, which is more constrained upon unrolling basic blocks, its predictions can generally be expected to be closer to measurements of \DefTPL.\looseness=-1

In Section~\ref{sec:predictor}, we develop two variants of our predictor, one for each notion of throughput.

\subsection{High-Level Pipeline Model}\label{sec:pipeline-model}

\pgfdeclarelayer{background}
\pgfdeclarelayer{background1}
\pgfdeclarelayer{foreground}
\pgfsetlayers{background,background1,main,foreground}

\tikzstyle{nodeStyle} = [draw, text width=6cm,  minimum height=1.75em, text centered]
\tikzstyle{port} = [draw, fill={rgb,255:red,135; green,220; blue,170}, text width=0.75cm, font=\fontsize{6}{7.2}\sffamily, text centered]
\tikzstyle{FU} = [draw, fill={rgb,255:red,212; green,170; blue,0}, text width=5.0em, font=\fontsize{6}{7.2}\sffamily, rotate=90, text centered]
\tikzstyle{arrow} = [draw, thick, color=black!80, font=\footnotesize\sffamily]

\newcommand{\background}[7]{%
  \begin{pgfonlayer}{background}
    \path (#1.west |- #2.north)+(-1,0.4) node (a1) {};
    \path (#3.east |- #4.south)+(+0.4,#5) node (a2) {};
    \path[fill=#6, draw=black!50]
      (a1) rectangle (a2);
    \path let \p{x}=(a1), \p{y}=($(a1)!0.5!(a2)$) in (\x{x}, \y{y})+(0.5,0) node (u1)[rotate=90]
      {#7};
  \end{pgfonlayer}}

\newcommand{\refsec}[1]{\large Section~\ref*{sec:#1}}

\begin{figure}[t]
\centering
\begin{tikzpicture}[scale=.7,transform shape,font=\fontsize{11}{13.2}\sffamily]
  \path node (nIC) [nodeStyle, fill={rgb,255:red,249; green,177; blue,166}] {Instruction Cache};
  \path (nIC.south)+(0.0,-0.75) node (nPreDec) [nodeStyle, text width=4cm, fill={rgb,255:red,171; green,204; blue,227}] {Predecoder};
  \path (nPreDec.south)+(0,-0.75) node (nIQ) [nodeStyle, text width=4cm, fill={rgb,255:red,198; green,233; blue,175}] {Instruction Queue (IQ)};
  \path (nIQ.south)+(-2.5,-0.75) node (nDSB) [nodeStyle, text width=1cm, fill={rgb,255:red,249; green,177; blue,166}] {DSB};
  \path (nIQ.south)+(0.0,-0.75) node (nDec) [nodeStyle, text width=2cm, fill={rgb,255:red,171; green,204; blue,227}] {Decoder};
  \path (nDec.south)+(0.0,-1.0) node [label={[xshift=-1.3em,text width=5cm]center:Instruction Decode Queue (IDQ)}] (nIDQ) [nodeStyle, fill={rgb,255:red,198; green,233; blue,175}, minimum height=1.2cm, minimum width=6.3cm] {};
  \path (nIDQ.east)+(-0.85,0.0) node (nLSD) [nodeStyle, text width=1cm, fill={rgb,255:red,171; green,204; blue,227}] {LSD};
  \path (nIDQ.south)+(0.0,-1) node (nRenamer) [nodeStyle, fill={rgb,255:red,171; green,204; blue,227}] {Renamer / Allocator};

  \path (nRenamer.south)+(0,-0.75) node (nReorder) [nodeStyle, text width=3cm, fill={rgb,255:red,198; green,233; blue,175}] {Reorder Buffer};
  \path (nReorder.south)+(0.0,-0.75) node (nRS) [nodeStyle, fill={rgb,255:red,135; green,220; blue,170}] {Scheduler};
  \path (nRS.south)+(-2.5,0.0) node[anchor=north] (nPort0) [port] {Port 0};
  \path (nRS.south)+(-1.5,0.0) node[anchor=north] (nPort1) [port] {Port 1};
  \path (nRS.south)+(-0.5,0.0) node[anchor=north] (nPort2) [port] {Port 2};
  \path (nRS.south)+(0.5,0.0) node[anchor=north] (nPort3) [port] {Port 3};
  \path (nRS.south)+(1.5,0.0) node[anchor=north] (nPort4) [port] {Port 4};
  \path (nRS.south)+(2.5,0.0) node[anchor=north] (nPort5) [port] {Port 5};

  \path (nPort0.south)+(0,-0.75) node[anchor=east] (nPort0FU) [FU] {ALU, V-MUL, \dots};
  \path (nPort1.south)+(0,-0.75) node[anchor=east] (nPort1FU) [FU] {ALU, V-ADD, \dots};
  \path (nPort2.south)+(0,-0.75) node[anchor=east] (nPort2FU) [FU] {Load, AGU};
  \path (nPort3.south)+(0,-0.75) node[anchor=east] (nPort3FU) [FU] {Load, AGU};
  \path (nPort4.south)+(0,-0.75) node[anchor=east] (nPort4FU) [FU] {Store Data};
  \path (nPort5.south)+(0,-0.75) node[anchor=east] (nPort5FU) [FU] {ALU, JMP, \dots};

  \begin{pgfonlayer}{background1}
    \path (nPort0FU.north |- nPort0FU.east)+(-0.25,0.25) node (ee_tl) {};
    \path (nPort5FU.south |- nPort5FU.west)+(+0.25,-0.25) node (ee_br) {};
    \path[fill={rgb,255:red,95; green,211; blue,188}, draw=black!50, rounded corners] (ee_tl) rectangle (ee_br);
  \end{pgfonlayer}


  \draw [->, arrow] (nIC.south) -- (nPreDec.north);
  \draw [->, arrow] (nPreDec.south) -- (nIQ.north);
  \draw [->, arrow] (nIQ.south) -- (nDec.north);
  \draw [->, arrow] (nDec.south) -- (nIDQ.north);
  \draw [->, arrow] (nDec.west) -- (nDSB.east);
  \draw [->, arrow] (nIDQ.south) -- (nRenamer.north);

  \draw [->, arrow] (nDSB.south) -- (nDSB.south |- nIDQ.north);

  \path (nRenamer.south west) -- (nRenamer.south) coordinate[pos=0.3] (p-Ren);
  \path (nRS.north west) -- (nRS.north) coordinate[pos=0.3] (p-RS);
  \draw [->, arrow] (nRenamer.south) -- (nReorder.north);
  \draw [->, arrow] (p-Ren)  -- (p-RS);

  \draw [->, arrow] (nReorder.east) -- node [above] {\textnormal retire} +(1.4,0) -- +(1.5,0);

  \draw [->, arrow, font=\fontsize{6}{7.2}\sffamily] (nPort0.south) -- node [right] {\textnormal\microop} +(-0,-0.5) -- (nPort0FU.east);
  \draw [->, arrow, font=\fontsize{6}{7.2}\sffamily] (nPort1.south) -- node [right] {\textnormal\microop} +(-0,-0.5) -- (nPort1FU.east);
  \draw [->, arrow, font=\fontsize{6}{7.2}\sffamily] (nPort2.south) -- node [right] {\textnormal\microop} +(-0,-0.5) -- (nPort2FU.east);
  \draw [->, arrow, font=\fontsize{6}{7.2}\sffamily] (nPort3.south) -- node [right] {\textnormal\microop} +(-0,-0.5) -- (nPort3FU.east);
  \draw [->, arrow, font=\fontsize{6}{7.2}\sffamily] (nPort4.south) -- node [right] {\textnormal\microop} +(-0,-0.5) -- (nPort4FU.east);
  \draw [->, arrow, font=\fontsize{6}{7.2}\sffamily] (nPort5.south) -- node [right] {\textnormal\microop} +(-0,-0.5) -- (nPort5FU.east);



  \path (3.25,0) node (nL2){};

  \background{nIC}{nIC}{nIDQ}{nIDQ}{-0.5}{{rgb,255:red,255; green,246; blue,213}}{\large ~\hspace{1.8cm}Front End}
  \background{nRS}{nRenamer}{nIDQ}{ee_br}{-0.25}{{rgb,255:red,213; green,255; blue,230}}{\large Back End\hspace{3cm}~}

  \path (nPreDec.east)+(2.8,0) node (predec) {\refsec{pipeline:predec}};

  \node at (predec |- nDec) (dec) {\refsec{pipeline:dec}};
  \node at (predec |- nLSD) (lsd) {\refsec{pipeline:lsd}};
  \node at (predec |- nRenamer) (issue) {\refsec{pipeline:issue}};

  \draw [dashed] (nPreDec.east) -- (predec.west);
  \draw [dashed] (nDec.east) -- (dec.west);
  \draw [dashed] (nLSD.east) -- (lsd.west);
  \draw [dashed] (nRenamer.east) -- (issue.west);

  \path (nDSB.west)+(-2.25,0) node (dsb) {\refsec{pipeline:dsb}};
  \node at (dsb |- nPort0) (ports) {\refsec{pipeline:ports}};

  \draw [dashed] (nDSB.west) -- (dsb.east);
  \draw [dashed] (nPort0.west) -- (ports.east);

  \path (nPreDec.east)+(1.6,0) node (predecleft) {};
  \path (nPort5FU.south)+(0,-1.2) node (portsSouth) {};

  \path (issue)+(0,-3.8) node (bla) {\refsec{pipeline:lat}};

  \draw [decorate, decoration = {brace}] (predecleft |- nRS.north) -- (predecleft |- portsSouth);

\end{tikzpicture}
\caption{Block diagram of pipeline model.} 
\label{pipeline}
\end{figure}
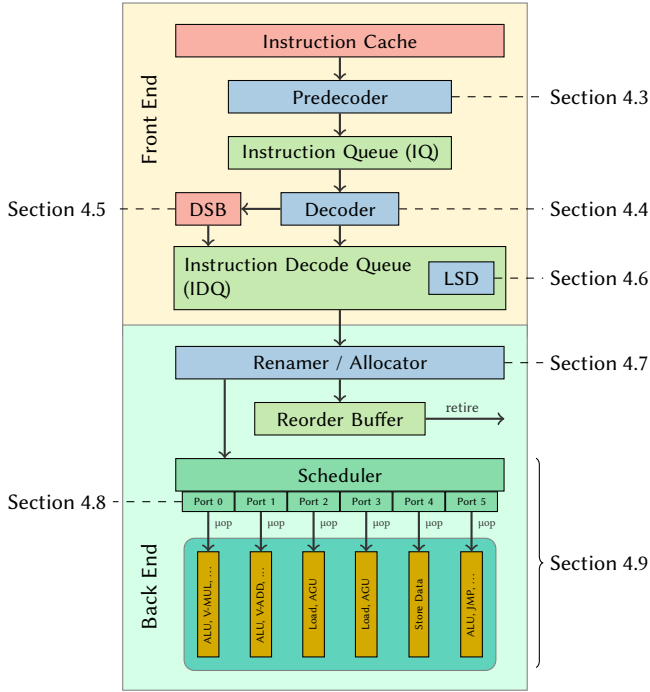

Our analytical predictor is based on the pipeline model depicted in Figure~\ref{pipeline}, which captures the high-level structure of recent Intel Core CPUs. 

The pipeline model consists of a front end and a back end.
The front end is responsible for fetching and decoding instructions into \microops,
while the back end, also known as the execution engine, is responsible for executing the \microops, respecting precedence constraints due to read-after-write dependencies, and eventually retiring them.

Both front end and back end can be further divided into multiple components:
\begin{itemize}
    \item The front end consists of the instruction cache (IC), the predecoder, the decoder, the decode stream buffer (DSB), and the loop stream detector (LSD).
    \item The back end consists of the renamer, which issues instructions to the scheduler, the scheduler, and the execution units, which are reached via the execution ports.\looseness=-1  \todo{later on we talk about the issue stage; should we introduce the term earlier?}
\end{itemize}
To enable different components to operate in parallel, buffers are placed between them.
For example, the decoder decodes input instructions and stores the resulting \microops in the instruction decode queue (IDQ).
The renamer fetches such \microops from the IDQ, performs its tasks and stores the \microops in the reorder buffer while also issuing them to the scheduler.
Each pipeline component can process multiple instructions or \microops in every cycle up to its parallel processing width.
As shown in prior work~\cite{Abel22}, the exact number of \microops processed by a component in a particular cycle depends on the properties of both the pipeline component and the \microops processed in that cycle.

Depending on the type of instruction and the state of the pipeline, instructions traverse different pipeline components.
In particular, instructions that are part of a loop will usually be processed by the predecoder and decoder only once, and will be served from the LSD or the DSB in subsequent iterations. 
Instructions that are not part of a loop, in contrast, are usually processed by the predecoder and the decoder.
Thus, steady-state throughput predictors for \DefTPU and \DefTPL will have to account for different subsets of the pipeline components.

Furthermore, instructions can be merged, split or eliminated at different stages of the pipeline.
For example, certain pairs of instructions can be merged in the Instruction Queue~(IQ).
These instructions are called ``macro fused'' instructions and they are treated as a single instruction by the rest of the pipeline.
Similarly, two \microops of some instruction types can be fused together in the decoding stage and are treated as a single \microop until they are split either by the renamer or when they enter the scheduler.
This is often referred to as ``micro fusion''.
Before reaching the splitting point, the decoded \microops are sometimes called \emph{fused-domain} \microops.
Finally, some \microops can be executed by the renamer itself and are not sent to the scheduler.

\subsection{Modeling Assumptions}\label{sec:assumptions}

The actual runtime performance of a basic block depends on a number of factors that are not captured by the pipeline model, some of which are actually impossible to know statically.
For example, memory accesses may be served from the L1 cache, the L2 cache, or the main memory, depending on the cache hierarchy and the state of the caches, which may in turn depend on the program's inputs.
Similarly, branches may be predicted correctly or incorrectly.
Floating-point operations can take a varying number of cycles depending on the operands, e.g., denormal operands may lead to a greater latency. 
Loads and stores may or may not alias.\looseness=-1

A common assumption in prior work is that basic-block throughput predictors are intended to analyze the performance of \emph{compute-bound} basic blocks.
In other words, the performance of the basic block is limited by the throughput of the front end or back end, and not by the throughput of the memory subsystem or by the branch predictor.
As a consequence, a common modeling assumption is that all memory accesses are served from the L1 cache and incur no TLB misses, unaligned loads, or bank conflicts.
Similarly, branch mispredictions are assumed to be negligible.
We adopt these assumptions in our analytical throughput model.
For a more detailed list of commonly unstated assumptions, we refer the reader to
~\cite{Abel22}.

\section{\toolname:\hspace{-0.125mm} An Analytical Throughput Model}\label{sec:predictor}

Our analytical basic-block throughput model \toolname is based on two hypotheses:
\begin{enumerate}
  \item The throughput of a basic block is determined by the slowest pipeline component or by dependency chains between instructions.
  \item Pipeline components can be analyzed independently from each other, as the pipeline stages are effectively decoupled from each other via buffers.
\end{enumerate}
Under these hypotheses, an accurate throughput predictor can be built from accurate predictors for individual pipeline components and by accounting for dependency chains.\looseness=-1

Furthermore, a throughput predictor obtained in this way is naturally interpretable, as it allows to easily identify the bottleneck component(s).
It also permits counterfactual reasoning, i.e., it allows to reason about the impact of improving the throughput of a particular pipeline component on the overall throughput.



\newcommand{\component}[1]{\texttt{#1}\xspace}
\newcommand{\predec}{\component{Predec}}
\newcommand{\dec}{\component{Dec}}
\newcommand{\issue}{\component{Issue}}
\newcommand{\pu}{\component{Ports}}
\newcommand{\lat}{\component{Precedence}}
\newcommand{\lsd}{\component{LSD}}
\newcommand{\dsb}{\component{DSB}}
\newcommand{\simpledec}{\component{SimpleDec}}
\newcommand{\simplepredec}{\component{SimplePredec}}

In the following sections, we will introduce throughput predictors for the following relevant pipeline components: 
\begin{itemize}
    \item \predec: throughput of the predecoder.
    \item \dec: throughput of the decoder.
    \item \issue: throughput of the issue stage.
    \item \pu: maximum throughput due to contention on execution ports.
\end{itemize}
For benchmarks that are executed as a loop, we also need to bound the throughput of the loop stream detector (\lsd) and the decode stream buffer (\dsb).
In addition, we capture the throughput bound due to precedence constraints among instructions in successive loop iterations (\lat).

A guiding principle in the design of the component predictors was to make the idealizing (and simplifying) assumption of optimal behavior rather than modeling intricate details of the actual behavior, where this is possible without sacrificing accuracy.

Before introducing all these component predictors, we show how their results can be combined to predict basic-block throughput.


\subsection{Throughput Prediction under Unrolling}

To predict the throughput of a basic block under unrolling, we determine the maximum throughput among all relevant components:
\begin{equation}\label{eq:tpu}
  \DefTPU = \max \{\predec, \dec, \issue, \pu, \lat\}
\end{equation}
A component is determined to be a \emph{bottleneck} if its throughput matches \DefTPU.
It is common for multiple components to be bottlenecks for a given benchmark.\looseness=-1

\subsection{Throughput Prediction for Loops}

\newcommand{\fe}{\component{FE}}

To predict the throughput of a basic block executed as a loop, we determine the maximum of the front-end throughput \fe and the throughput of the back-end components:
\begin{equation}\label{eq:tpl}
  \DefTPL = \max \{\fe, \issue, \pu, \lat\}
\end{equation}

The front-end throughput depends on whether the benchmark is affected by the JCC erratum\footnote{As a mitigation to the ``Jump Conditional Code'' (JCC) erratum, Skylake CPUs do not cache blocks that contain a jump instruction that crosses or ends on a 32-byte boundary~\cite{jccErratum}.}.
In that case, neither the LSD nor the DSB are used and the basic block needs to go through the predecoder and the decoder.
If the LSD is enabled for the microarchitecture-under-analysis\footnote{On Skylake CPUs it was disabled due to the SKL150 erratum~\cite{sklSpecUpdate}.} and the benchmark fits into the instruction decode queue, the throughput is determined by \lsd.
Otherwise, the benchmark is serviced from the DSB.

\begin{equation}\label{eq:fe}
  \fe = \begin{cases}
\max \{ \predec, \dec \} & \textit{if benchmark is affected } \\
				   & ~~~\textit{by the JCC erratum}\\
\lsd & \textit{else if LSD is enabled and }\\
	& ~~~\textit{\#\microops $\leq$ IDQWidth}\\
\dsb & \textit{else}\\
\end{cases}
\end{equation}

\subsection{Predecoder}\label{sec:pipeline:predec}

The predecoder fetches instructions from the instruction cache in aligned 16-byte blocks.
It detects the beginning of each instruction in these 16-byte blocks and stores the predecoded instructions into the instruction queue (IQ).
This is necessary as each instruction can be between 1 to 15 bytes long and finding the beginning of a subsequent instruction may require an inspection of several bytes of the current instruction.

To determine the steady-state behavior of the predecoder, we first determine the number of loop iterations~$u$ after which the behavior of the predecoder will start repeating.
Let $l$ be the length of the benchmark in bytes.
As the predecoder considers blocks of size 16 bytes, the number of iterations $u$ is determined by
$$u =
\begin{cases}
\frac{lcm(l, 16)}{l} & \textit{for unrolling,}\\
1 & \textit{for loops.}\\
\end{cases}$$

Then the throughput of the predecoder can be obtained by computing the number of cycles required to predecode $u$~unrolled copies of the basic block divided by $u$.

The predecoder can predecode up to five instructions per cycle.
If there are more than five instructions in a 16-byte block, the following up to five instructions in the same block are predecoded in the next cycle, and so on. 
It was shown in prior work~\cite{Abel22} that when an instruction crosses the 16-byte boundary, depending on whether the nominal opcode of the instruction lies in the first 16-byte block or not, there may be a one cycle penalty in its predecoding.
Further, if there are instructions with a length-changing prefix (LCP), the predecoder has to use a special algorithm that results in a penalty of three cycles for every such instruction.\looseness=-1

The number of 16-byte blocks in a benchmark of length~$l$ that is unrolled $u$ times is $n = \frac{u \cdot l}{16}$.
Let $L(b)$ be the number of instruction instances whose last byte is in the $b$-th 16-byte block.
Let $O(b)$ be the number of instruction instances whose first byte of the nominal opcode (i.e., the first byte that does not belong to a prefix) is in the $b$-th 16-byte block but whose last byte is not in the same block.
Then the number of cycles required to predecode all the non-LCP instructions in block $b$ is given by

$$ cycle_{NLCP}(b) = \left\lceil \frac{L(b)+O(b)}{5}\right\rceil.$$

Next we add the extra cycles needed due to the penalty for LCP instructions.
Let $LCP(b)$ be the number of instruction instances whose first byte of the nominal opcode is in the $b$-th 16-byte block, and which have a length-changing prefix.
Also, let $L(-1) = L(n-1)$, and $O(-1) = O(n-1)$.
The number of cycles required to predecode the LCP instructions in block $b$ is given by
$$ cycle_{LCP}(b) = \max \left(0, 3 \cdot LCP(b) - \left(cycle_{NLCP}(b-1) - 1\right) \right). $$

We assume that after the first cycle of predecoding block $b-1$, the predecoder fetches block $b$ and the length-decoding algorithm starts processing the LCP instructions.
Hence, we deduct all but one cycle of predecoding block $b-1$ from the penalty of predecoding LCP instructions truncating the difference to zero.
Putting everything together, the throughput of the predecoder is predicted as
$$ \predec = \frac{\sum_{b=0}^{n-1} \left( cycle_{NLCP}(b) + cycle_{LCP}(b) \right)}{u}.$$

For comparison, we also consider a simpler model for the throughput of the predecoder.
In this simpler model, we assume that the predecoder predecodes a 16-byte block in each cycle.
Hence, for a benchmark of length $l$ bytes, the simple predecoder throughput is given by

$$\simplepredec = \frac{l}{16}.$$

\subsection{Decoder}\label{sec:pipeline:dec}

The decoding unit decodes up to four instructions per cycle and stores the decoded \microops into the instruction decode queue (IDQ).
The decoding unit consists of three simple decoders and one complex decoder.
The complex decoder can decode instructions with up to four \microops whereas the simple decoders can only handle instructions with one \microop.
However, the complex decoder always decodes the first fetched instruction in a cycle.
The decoding unit assigns an incoming instruction to a decoder based on the current state of the decoding unit and the properties of the incoming instruction.\looseness=-1

\LinesNumbered
\newcommand{\assign}{\leftarrow}
\newcommand{\curDec}{\textit{curDec}}
\newcommand{\nAvailableSimpleDecoders}{\textit{nAvailableSimpleDecoders}}
\newcommand{\nComplexDecInRound}{\textit{nComplexDecInIteration}}
\newcommand{\firstInstrOnDecInRound}{\textit{firstInstrOnDecInIteration}}
\newcommand{\round}{\textit{iteration}}
\newcommand{\instructions}{\textit{instructions}}
\newcommand{\cycles}{\textit{cycles}}

\begin{algorithm}[t!]
\caption{Decoder}\label{decAlg}
\DontPrintSemicolon
$\curDec \assign \#decoders - 1$\;
$\nAvailableSimpleDecoders \assign 0$\;
$\nComplexDecInRound \assign [0, \dots, 0]$\;
$\firstInstrOnDecInRound \assign [-1, \dots, -1]$\;
$\round \assign 0$\;
\While{True}{\label{line:start}
	$\round \assign \round + 1$\;
	$\nComplexDecInRound[\round] \assign 0$\;
	\For{$i \in \instructions$}{
		\If{i \text{requires the complex decoder}} {
			$\curDec \assign 0$\;
			$\nAvailableSimpleDecoders \assign i.\nAvailableSimpleDecoders$ \tcp{as provided by uops.info}
		}\Else{
    		\If{$(\nAvailableSimpleDecoders =~0) \vee ((\curDec + 1 =~ \#decoders - 1) \wedge (\text{i is macro-fusible}) \wedge (\text{microarch. cannot decode macro-fusible}$
			$\text{instr. on last decoder}))$} {
    			$\curDec \assign 0$\;
    			$\nAvailableSimpleDecoders \assign \#decoders - 1$\;
    		}\Else{
    			$\curDec \assign \curDec + 1$\;
    			$\nAvailableSimpleDecoders \assign \nAvailableSimpleDecoders - 1$\;
    		}
		}
		\If{i \text{is branch instruction}} {
			$\nAvailableSimpleDecoders \assign 0$\;
		}
		\If{$curDec = 0$} {
			$\nComplexDecInRound[\round] \assign \nComplexDecInRound[\round] + 1$\;
		}
		\If{i \text{is the first instruction in the benchmark}}{
			$f \assign \firstInstrOnDecInRound[\curDec]$\;
			\If{$f \geq 0$}{\label{line:secondtime}
        $u \assign \round - f$\;
        $\cycles(u) \assign \sum_{r=f}^{\round-1} \nComplexDecInRound[r]$\;
				\Return{$\frac{\cycles(u)}{u}$}\;
			}\Else{
				$\firstInstrOnDecInRound[curDec] \assign \round$\;
      }\label{line:end}
		}
	}
}
\negvspace
\end{algorithm}


Our approach to predict the throughput of the decoding unit is to simulate the allocation of instructions to decoders until the first instruction of the benchmark is allocated to the same decoder for the second time.
At that point, the decoder has reached its steady-state behavior and we can compute the throughput of the decoding unit based on its timing up to that point.

Let $u$ be the number of benchmark iterations between the first and second allocation of the first instruction of the benchmark to the same decoder.
Let $cycles(u)$ be the time between these two allocations.
Then the throughput of the decoding unit is given by
$$\dec = \frac{\cycles(u)}{u}.$$

Algorithm~\ref{decAlg} illustrates our approach to computing the throughput of the decoding unit in more detail.
The array $\nComplexDecInRound$ stores the number of times the complex decoder is used to decode an iteration of the basic block.
This corresponds to the number of cycles needed to decode the basic block in that particular iteration.
The array $\firstInstrOnDecInRound$ tracks the decoder that the first instruction in the basic block is allocated to in each iteration.
The algorithm iterates over all instructions in the basic block repeatedly and allocates decoders to them (lines \ref{line:start} to \ref{line:end}) until the first instruction of the benchmark is allocated to the same decoder for the second time (line \ref{line:secondtime}).
Then $f = \firstInstrOnDecInRound[\curDec]$ is the iteration in which the first instruction of the benchmark was first allocated to the current decoder $\curDec$.
The number of basic-block iterations between these two allocations is $u = \round - f$ and the time between these two allocations is $cycles(u) = \sum_{r=f}^{\round-1} \nComplexDecInRound[r]$.

Similar to the predecoder, we also consider a simpler model for the decoder for comparison.
Let $n$ be the number of instructions of the benchmark (ignoring instructions that are macro fused with the preceding instruction).
Let~$c$ be the number of instructions that require the complex decoder.
Let~$d$ be the number of decoders in the microarchitecture.
Then \simpledec is given by
$$\simpledec = \max \left\{\frac{n}{d}, c\right\}.$$

\subsection{Decoded Stream Buffer (DSB)}\label{sec:pipeline:dsb}

The decoded stream buffer, also called the \microop cache, stores decoded \microops. 
For loops that are bottlenecked by decoding, it can improve the throughput by storing \microops that are decoded in the first iteration of the loop, forwarding them to the renamer in the subsequent iterations.

We predict the throughput of the DSB simply as the ratio of the number of \microops in the benchmark to the maximum number of \microops that can be forwarded from the DSB to the renamer in a single cycle.
Let $n$ be the number of (fused-domain) \microops of the benchmark.
Let $l$ be the length (in bytes) of the benchmark.
Let $w$ be the width of the DSB.
The throughput of the DSB is then given by
$$ \dsb =
\begin{cases}
\left\lceil \frac{n}{w} \right\rceil & l < 32,\\
\frac{n}{w} & l \geq 32.
\end{cases}$$
The first case captures that, after a branch instruction, the CPU cannot load additional \microops in the same cycle if they are in the same 32-byte block as the branch instruction.

\subsection{Loop Stream Detector (LSD)}\label{sec:pipeline:lsd}
The loop stream detector detects if a loop's \microops fit into the IDQ completely.
In such situations, it locks the \microops in the IDQ and continuously streams them to the renamer without waiting for the DSB or the decoders.
However, the last \microop of the current loop iteration and the first \microop of the following iteration cannot be streamed in the same cycle.
For small loops, this can be a significant bottleneck as the number of \microops streamed per cycle could be less than the issue width of the microarchitecture.
In such cases, the LSD unrolls the loop to increase the number of \microops streamed per cycle.
The number of times the LSD unrolls the loop was reverse engineered in~\cite{Abel22}.

Let $n$ be the number of (fused-domain) \microops of the benchmark.
Let $i$ be the issue width of the microarchitecture.
Let $u$ be the number of times the LSD unrolls the benchmark, as provided at~\footnote{\url{https://github.com/andreas-abel/uiCA/blob/master/microArchConfigs.py}}.
The throughput of the LSD is then given by\looseness=-1
$$\lsd = \frac{\left\lceil \frac{n \cdot u}{i}\right\rceil}{u}.$$

\subsection{Issue}\label{sec:pipeline:issue}

As mentioned in Section~\ref{sec:pipeline-model}, the renamer fetches \microops from the IDQ and issues them to the scheduler.
It is also responsible for allocating resources for loads and stores and assigning execution ports to \microops.
The number of \microops that can be issued per cycle is bound by the \emph{issue width} of the microarchitecture.

Micro-fused \microops that are split by the renamer are called \emph{unlaminated} \microops after being split.
They are issued as two \microops to the scheduler.

We model the throughput of the renamer as the ratio of the number of \microops issued by the renamer to the issue width of the microarchitecture.
Let $n$ be the number of (fused-domain, but after unlamination) \microops of the benchmark.\todo{not so important now, but i'm confused about the interpretation of ``fused-domain, but after unlamination'' having read the description earlier in the paper.}
Let~$i$ be the issue width of the corresponding microarchitecture.
The throughput is then given by
$$\issue = \frac{n}{i}.$$
Note that \lsd dominates \issue in \DefTPL if the LSD is active, but \issue may be the bottleneck in other cases.


\subsection{Execution Ports}\label{sec:pipeline:ports}
%

Each \microop can be dispatched to a subset of the pipeline's execution ports.
Contention on execution ports may slow down the execution of a basic block.
Which ports a \microop can be dispatched to has been reverse engineered in \cite{Abel19}.
The renamer assigns each \microop to one of the possible execution ports, attempting to evenly distribute \microops across ports.
A precise model of the port-assignment algorithm employed by renamers in recent Intel microarchitectures has been described in \cite{Abel22}.
In this work, we simply assume that the renamer optimally distributes the load, which is close to reality in most cases.
Under this idealizing assumption, \cite{Abel19} previously introduced a linear program to determine the throughput bound due to port contention.

Here, we propose a simpler, more efficient approach that leads to the same bound on all of the BHive benchmarks.
%
%
Our approach is based on the following observation:
If a benchmark contains a subset of \microops of size $u$ that can collectively be dispatched to port combination $pc$ (i.e., the set of ports that the $u$ \microops may be dispatched to), then the benchmark's throughput is limited by $\frac{u}{|pc|}$, as each port accepts at most one \microop per cycle.

Rather than considering every port combination that some subset of a benchmark's \microops can be dispatched to, we heuristically consider only port combinations required by pairs of \microops.
Let $PC$ be the set of all port combinations that are used by the \microops of a benchmark, and let $PC' = \{pc \cup pc' \mid pc, pc' \in PC\}$ be the set of port combinations of pairs of \microops in the benchmark. 
Then we predict the throughput bound \pu due to port contention as follows:

\begin{align*}
   \pu = \max_{pc \in PC'} \left\{ \frac{u}{\vert pc \vert} ~\middle|~ u = \textit{number of \microops in benchmark} \right.\\
    \left.\phantom{\frac{u}{\vert pc \vert}}\textit{whose port combination is a subset of pc}\right\}
\end{align*}


We exclude \microops from instructions that may be eliminated or that are macro-fused with preceding instructions.
To provide interpretable feedback in case \pu is a bottleneck component, we can extract the instructions whose \microops experience the maximal port contention. \todo{i guess the implementation does not allow to do this yet, but it would be easy to add and contribute to interpretability.}

\subsection{Precedence Constraints}\label{sec:pipeline:lat}

In the steady state, the throughput of a basic block may be limited by precedence constraints of instructions across basic-block iterations due to read-after-write dependencies.

To determine the bound on the throughput due to such constraints, we construct a weighted directed dependence graph.
The nodes of this graph correspond to the values consumed and produced by the instructions of the benchmark.
For each instruction, we create a node for each value it consumes and each value it produces.
These nodes are then connected by edges that capture the latency between the consumption of the source value and the production of the target value.
We obtain the necessary data for this construction from uops.info~\cite{Abel19}.

We add $0$-latency edges between produced and consumed values to capture the dependencies between instructions of the benchmark.
This includes dependencies within a single iteration of the basic block as well as between consecutive iterations of the basic block.
Such dependency edges are associated with an iteration count.
This iteration count is $0$ for intra-iteration dependencies and $1$ for inter-iteration dependencies.
The iteration count of all edges capturing instruction latencies is $0$.

Now consider a cycle through such a dependence graph.
The sum of the latency weights of the edges on this cycle is the latency of the cycle.
The sum of the iteration counts of the edges on this cycle is the number of iterations the cycle spans.
The throughput bound due this particular cycle is the ratio of the latency to the number of iterations.
The throughput bound of the benchmark as a whole is the maximum such ratio across all cycles through the graph.
Note that by construction any cycle must include at least one inter-iteration edge and so the iteration count is at least one.
This ratio can be obtained using optimum cycle ratio algorithms~\cite{Dasdan04}.
Our implementation employs Howard's value iteration algorithm~\cite{Howard60,Dasdan04} to compute it.

To provide interpretable feedback in case \lat is a bottleneck component, we can extract the dependency chain that exhibits the maximal latency.

Our approach is reminiscent of similar analyses in the context of \emph{modulo scheduling}~\cite{Lam88,Huff93,Rau94}, where precedence constraints induce a \emph{recurrence-constrained minimum initiation interval} between successive loop iterations, which corresponds to the throughput bound we compute here.
Modulo scheduling also accounts for resource constraints, which correspond to the throughput bounds induced by the various pipeline components in our approach and discussed in previous sections.





\section{Implementation}\label{sec:implementation}

\begin{figure}[t]
  \begin{center}
\begin{tikzpicture}[
    node distance=2cm,
    process/.style={rectangle, draw, fill=green!20, text centered, minimum width=2cm, minimum height=1cm},
    datasource/.style={rectangle, draw, fill=blue!20, text centered, rounded corners, minimum width=2cm, minimum height=1cm},
    result/.style={rectangle, draw, fill=red!20, text centered, minimum width=2cm, minimum height=1cm}
]
  \node [datasource, align=center] (input) {Basic Block};

  \node [process, right of=input, xshift=1cm, align=center] (predictor) {\Large \toolname};
  \node [datasource, align=center, above left = 0.7 cm and -0.7 cm of predictor] (data1) {Instr. data from\\ uops.info};
  \node [datasource, align=center, right of=data1, xshift=1cm] (data2) {\textmu Arch. \\ Configuration};
  \node [process, below of=predictor, xshift=0cm, yshift=0.3cm, align=center] (disassembler) {XED \\ Disassembler};

  \node [result, right of=predictor, xshift=1cm, align=center] (result) {Predicted \\ Throughput};

  \draw[->,arrow] (data1) -- (predictor);
  \draw[->,arrow] (data2) -- (predictor);
  \draw[->,arrow] (input) -- (predictor);
  \draw[->,arrow] (predictor) -- (result);
  \draw[dashed,->, arrow] (disassembler) -- (predictor);
\end{tikzpicture}
\end{center}
\caption{Flow diagram of \toolname.\label{fig:flow-diagram}}
\end{figure}
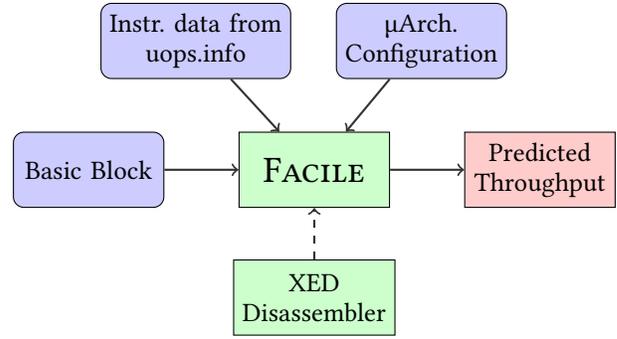


\toolname is implemented in Python 3.
It takes a basic block binary as input and returns its predicted throughput.
To this end, the binary is first disassembled using a modified version of the Intel X86 Encoder Decoder (XED) library, which provides a Python interface and is provided at~\footnote{\url{https://github.com/andreas-abel/XED-to-XML}}.
Then, the throughput of the basic block is predicted using the model described in Section~\ref{sec:predictor}.
The throughput depends both on high-level parameters of the microarchitecture, such as the issue width, the DSB width, or the number of decoders, as well as on properties of individual instructions, such as their port mapping and latency.
For this information, \toolname relies on two sources:
\begin{enumerate}
  \item Microarchitecture-specific parameters are obtained from the microarchitecture configuration files from uiCA~\cite{uica}.
  \item Instruction-specific parameters are obtained from the uops.info database~\cite{Abel19,uops-instructions}.
\end{enumerate}
The basic flow of \toolname is depicted in Figure~\ref{fig:flow-diagram}.
\toolname can be invoked from the command line or by calling the predictor function from other Python scripts, which is how we use it in our experiments.

\section{Experimental Evaluation}\label{sec:evaluation}

In this section, we experimentally evaluate three different aspects of \toolname:

\begin{compactitem}
  \item \textbf{Accuracy:} We compare the accuracy of \toolname to that of other throughput predictors and analyze the impact of the various components on its accuracy.
  \item \textbf{Computational efficiency:} We analyze the efficiency of \toolname by comparing it to that of other throughput predictors as well as measuring the contribution of individual components to the execution time of \toolname.\todo{terminology: efficiency vs performance vs speed; also predictors vs models}
  \item \textbf{Interpretability:} We exploit its interpretability to gain insights into the evolution of Intel microarchitectures.
\end{compactitem}

\subsection{Benchmarks}

The BHive~\cite{Chen19} benchmark suite is often used for evaluating throughput predictors for basic blocks.
It is an open-source benchmark suite with more than 300,000 basic blocks that have been extracted from applications from different domains such as numerical computing, databases, compilers, machine learning and cryptography.
However, as pointed out by prior work~\cite{Abel22}, there are two important issues with the BHive benchmark suite.
First, there are basic blocks in the BHive suite that do not adhere to the modeling assumptions discussed in Section~\ref{sec:assumptions}.
Second, the benchmarks in the BHive suite do not end in branch instructions.
Hence, the measurements made by the BHive profiler are based on the \DefTPU notion of throughput while most previous predictors use the \DefTPL notion of throughput.

To address these issues, we use a modified version of the BHive benchmarks as obtained from~\cite{uica-eval}.
It contains a subset of the BHive benchmarks that conforms to the modeling assumptions from Section~\ref{sec:assumptions}.
Furthermore, it also contains variants of the benchmarks from the BHive suite that end in a branch instruction.
This allows for a more meaningful comparison with predictors that are based on the \DefTPL notion of throughput.

We refer to the benchmarks based on the \DefTPU notion of throughput as \bhiveu and those based on the \DefTPL notion as \bhivel in the remainder of this work.

\begin{table}[t]
  \caption{Microarchitectures used for the evaluation.}
  \negvspace
    \label{tab:microarchitectures}
    \begin{center}
    \begin{tabular}{lccl}
    \toprule
    \textbf{{$\mu$}Arch} & \textbf{Abbr.} & \textbf{Released} & \textbf{CPU} \\
    \midrule
    Rocket Lake & RKL & 2021 & Intel Core~i9-11900 \\
    Tiger Lake & TGL & 2020 & Intel Core~i7-1165G7 \\
    Ice Lake & ICL & 2019 & Intel Core~i5-1035G1 \\
    Cascade Lake & CLX & 2019 & Intel Core i9-10980XE \\
    Skylake & SKL & 2015 & Intel Core~i7-6500U \\
    Broadwell & BDW & 2015 & Intel Core~i5-5200U \\
    Haswell & HSW & 2013 & Intel Xeon~E3-1225~v3 \\
    Ivy Bridge & IVB & 2012 & Intel Core~i5-3470 \\
    Sandy Bridge & SNB & 2011 & Intel Core~i7-2600\\
    \bottomrule
    \end{tabular}
    \end{center}
  \end{table}

\subsection{Accuracy}

In this section, we study the accuracy of \toolname.
We compare its accuracy to that of other throughput predictors on all modern Intel Core microarchitectures listed in Table~\ref{tab:microarchitectures}.

As prior work~\cite{Abel22,Chen19,Renda20}, we use the following metrics to compare the accuracy of the throughput predictors:
\begin{compactitem}
  \item Mean Absolute Percentage Error (MAPE): The mean absolute percentage error of the predictions made by the throughput predictor, relative to the measured throughputs.
  For a set of pairs of measured and predicted throughput values, $S = \{(m_i, p_i)\}_{i=1}^n$, the MAPE is defined as:
  $$
      MAPE(S) = \frac{1}{n} \sum_{i=1}^n \left\vert \frac{m_i - p_i}{m_i} \right\vert
  $$

  \item Kendall's tau coefficient~\cite{Kendall38}: Kendall's tau is a measure of the correlation between two rankings.
  We use Kendall's tau to compare pairs of benchmarks based on their throughput.
  As mentioned by prior works~\cite{Abel22,Chen19}, for certain applications, the relative ordering of different basic blocks based on their throughput is more important than the absolute throughput values.
\end{compactitem}

For our comparison, we use the most recent version of the throughput predictors wherever possible.
We use OSACA~\cite{Laukemann18} at version 0.5.0, llvm-mca~\cite{llvmmca} at version 15.0.7 and CQA~\cite{maqao} at version 2.17.0.
Additionally, since DiffTune is based on llvm-mca 8.0.1, we also include a comparison with this version of llvm-mca.
For DiffTune~\cite{Renda20}, we use the version at~\cite{diffTune} with the models from the paper that are available at~\footnote{\url{https://github.com/ithemal/DiffTune/issues/1}}. 
``learning-bl'' corresponds to the baseline for evaluating llvm-mca parameter learning approaches proposed in~\cite{Abel22b}; we use the code and models that are available at~\cite{diffTune-revisited}.
We use IACA~\cite{iacaGuide} at versions 3.0 and 2.3 as version 3.0 does not support the older microarchitectures.
We use Ithemal~\cite{mendis19a} at \cite{ithemal} and uiCA~\cite{Abel22} at \cite{uica}.
We could not experimentally evaluate GRANITE~\cite{sykora22} as the trained model is not publicly available.
We execute all the throughput predictors with a timeout of one hour.
If a throughput predictor crashes or does not terminate within the timeout, we report a throughput of zero for that benchmark.
We round the predictions to two decimal digits as the measurements were rounded in the same way.
\looseness=-1

\begin{table}
\caption{Comparison of predictors on \bhiveu and \bhivel.}
\label{tab:accuracy-results}
\begin{center}
\begin{tabular}{clrcrc}
\toprule
& & \multicolumn{2}{c}{\textbf{\bhiveu}} & \multicolumn{2}{c}{\textbf{\bhivel}}\\  \cmidrule(lr){3-4}\cmidrule(lr){5-6}
\textbf{{$\mu$}Arch}  & \textbf{Predictor} & \textbf{MAPE} & \textbf{Kendall} & \textbf{MAPE} & \textbf{Kendall} \\
\midrule
\multirow{4}{*}{RKL}  & \toolname & 0.42\% & 0.9860 & 1.04\% & 0.9731\\
                      & \uiCA & 0.49\% & 0.9835 & 0.92\% & 0.9755 \\
                      & llvm-mca-15 & \wrongDef{25.84\%} & \wrongDef{0.6553} & 15.20\% & 0.8380\\
                      & CQA 2.17.0 & \wrongDef{19.18\%}  & \wrongDef{0.7085} & 5.69\% & 0.9123 \\
\midrule
\multirow{4}{*}{TGL}  & \toolname & 1.15\% & 0.9717 & 1.62\% & 0.9617\\
                      & \uiCA & 0.97\% & 0.9769 & 0.98\% & 0.9731 \\
                      & llvm-mca-15 & \wrongDef{25.92\%} & \wrongDef{0.6989} & 13.98\% & 0.8389\\
                      & CQA 2.17.0 & \wrongDef{21.76\%} & \wrongDef{0.6996} & 5.44\% & 0.9139 \\
\midrule
\multirow{5}{*}{ICL}  & \toolname & 1.17\% & 0.9713 & 1.36\% & 0.9681\\
                      & \uiCA & 1.00\% & 0.9771 & 0.77\% & 0.9759 \\
                      & OSACA & \wrongDef{20.11\%} & \wrongDef{0.7032} & 10.07\% & 0.7865 \\
                      & llvm-mca-15 & \wrongDef{25.55\%} & \wrongDef{0.6970} & 13.64\% & 0.8512 \\
                      & CQA 2.17.0 & \wrongDef{21.82\%} & \wrongDef{0.6982} & 5.03\% & 0.9256 \\
\midrule
\multirow{4}{*}{CLX}  & \toolname & 0.62\% & 0.9692 & 1.24\% & 0.9720\\
                      & \uiCA & 0.45\% & 0.9713 & 0.65\% & 0.9825 \\
                      & llvm-mca-15 & \wrongDef{23.17\%} & \wrongDef{0.7211} & 13.21\% & 0.8060 \\
                      & OSACA & \wrongDef{19.02\%} & \wrongDef{0.7910} & 10.61\% & 0.8350 \\
\midrule
\multirow{11}{*}{SKL} & \toolname & 0.69\% & 0.9753 & 1.08\% & 0.9752\\
                      & \uiCA & 0.45\% & 0.9798 & 0.38\% & 0.9895 \\
                      & Ithemal & 8.28\% & 0.8172 & \wrongDef{13.66\%} & \wrongDef{0.7582} \\
                      & IACA 3.0 & \wrongDef{13.49\%} & \wrongDef{0.7802} & 14.26\% & 0.8290 \\
                      & IACA 2.3 & \wrongDef{11.85\%} & \wrongDef{0.8071} & 8.42\% & 0.8477 \\
                      & OSACA & \wrongDef{14.04\%} & \wrongDef{0.7810} & 10.04\% & 0.8223 \\
                      & llvm-mca-15 & \wrongDef{15.61\%} & \wrongDef{0.7258} & 12.01\% & 0.8015 \\
                      & llvm-mca-8 & \wrongDef{15.39\%} & \wrongDef{0.7434} & 11.98\% & 0.8021 \\
                      & DiffTune & 24.48\% & 0.6626 & \wrongDef{104.88\%} & \wrongDef{0.6426} \\
                      & learning-bl & 12.02\% & 0.7933 & 15.63\% & 0.8020 \\
                      & CQA 2.17.0 & \wrongDef{16.07\%} & \wrongDef{0.7375} & 6.58\% & 0.8972 \\
\midrule
\multirow{7}{*}{BDW}  & \toolname & 0.73\% & 0.9852 & 1.44\% & 0.9704\\
                      & \uiCA & 1.08\% & 0.9805 & 0.60\% & 0.9841 \\
                      & IACA 3.0 & \wrongDef{14.69\%} & \wrongDef{0.8012} & 11.47\% & 0.8725 \\
                      & IACA 2.3 & \wrongDef{13.22\%} & \wrongDef{0.8206} & 5.84\% & 0.8928 \\
                      & OSACA & \wrongDef{17.00\%} & \wrongDef{0.7621} & 8.95\% & 0.8528 \\
                      & llvm-mca-15 & \wrongDef{14.23\%} & \wrongDef{0.7793} & 16.71\% & 0.8286 \\
                      & CQA 2.17.0 & \wrongDef{16.11\%} & \wrongDef{0.7593} & 5.00\% & 0.9222 \\
\midrule
\multirow{11}{*}{HSW} & \toolname & 1.26\% & 0.9791 & 1.50\% & 0.9690\\
                      & \uiCA & 0.76\% & 0.9850 & 0.59\% & 0.9842 \\
                      & Ithemal & 7.38\% & 0.8400 & \wrongDef{16.19\%} & \wrongDef{0.7700} \\
                      & IACA 3.0 & \wrongDef{15.04\%} & \wrongDef{0.8080} & 12.00\% & 0.8733 \\
                      & IACA 2.3 & \wrongDef{13.13\%} & \wrongDef{0.8291} & 5.79\% & 0.8925 \\
                      & OSACA & \wrongDef{17.49\%} & \wrongDef{0.7644} & 9.06\% & 0.8517 \\
                      & llvm-mca-15 & \wrongDef{20.29\%} & \wrongDef{0.7835} & 18.97\% & 0.8259 \\
                      & llvm-mca-8 & \wrongDef{21.08\%} & \wrongDef{0.7784} & 19.46\% & 0.8171 \\
                      & DiffTune & 24.80\% & 0.6997 & \wrongDef{138.47\%} & \wrongDef{0.6925} \\
                      & learning-bl & 11.82\% & 0.8302 & 19.26\% & 0.8052 \\
                      & CQA 2.17.0 & \wrongDef{16.23\%} & \wrongDef{0.7668} & 5.05\% & 0.9229 \\
\midrule
\multirow{10}{*}{IVB}  & \toolname & 1.80\% & 0.9568 & 1.52\% & 0.9624\\
                      & \uiCA & 1.50\% & 0.9609 & 1.11\% & 0.9495 \\
                      & Ithemal & 7.08\% & 0.8212 & \wrongDef{12.43\%} & \wrongDef{0.7785} \\
                      & IACA 2.3 & \wrongDef{13.94\%} & \wrongDef{0.7739} & 11.54\% & 0.8271 \\
                      & OSACA & \wrongDef{23.87\%} & \wrongDef{0.6388} & 15.80\% & 0.7467 \\
                      & llvm-mca-15 & \wrongDef{22.79\%} & \wrongDef{0.7656} & 20.76\% & 0.8154 \\
                      & llvm-mca-8 & \wrongDef{22.93\%} & \wrongDef{0.7622} & 20.76\% & 0.8138 \\
                      & DiffTune & 26.21\% & 0.6470 & \wrongDef{82.94\%} & \wrongDef{0.7516} \\
                      & learning-bl & 13.20\% & 0.7654 & 16.89\% & 0.7937 \\
                      & CQA 2.17.0 & \wrongDef{40.60\%} & \wrongDef{0.5167} & 4.05\% & 0.9174 \\
\midrule
\multirow{6}{*}{SNB}  & \toolname & 1.95\% & 0.9586 & 1.33\% & 0.9742\\
                      & \uiCA & 1.91\% & 0.9613 & 0.98\% & 0.9650 \\
                      & IACA 2.3 & \wrongDef{11.91\%} & \wrongDef{0.8194} & 9.95\% & 0.8482 \\
                      & OSACA & \wrongDef{12.63\%} & \wrongDef{0.7939} & 13.08\% & 0.8137 \\
                      & llvm-mca-15 & \wrongDef{22.67\%} & \wrongDef{0.8069} & 18.34\% & 0.8455 \\
                      & CQA 2.17.0 & \wrongDef{32.94\%} & \wrongDef{0.6274} & 4.07\% & 0.9238 \\
\bottomrule
\end{tabular}
\end{center}
\end{table}

\subsubsection*{\toolname versus other predictors}
Table~\ref{tab:accuracy-results} shows the MAPE and Kendall's tau for our experiments on the \bhiveu and \bhivel benchmarks.
For completeness, we evaluate all throughput predictors on both \bhiveu and \bhivel benchmarks.
For predictors based on the \DefTPL notion of throughput, we present the \bhiveu results in gray and vice versa for the predictors based on \DefTPU.

The main finding is that \toolname performs similarly to or slightly worse than the state-of-the-art uiCA on all microarchitectures, but significantly better than all other throughput predictors.
We also note that the gap between \toolname and uiCA is generally smaller on \bhiveu than on \bhivel.\looseness=-1

Although we were unable to run GRANITE ourselves, based on the results reported in the paper~\cite{sykora22} for the BHive benchmark suite, we note that \toolname significantly outperforms GRANITE on all microarchitectures.
\todo{In general, \toolname performs better on \bhiveu benchmarks. Can we add some explanation for this?}
\todo{Do we have an intuition as to why uiCA is still a bit more accurate in most cases?}

\begin{figure*}
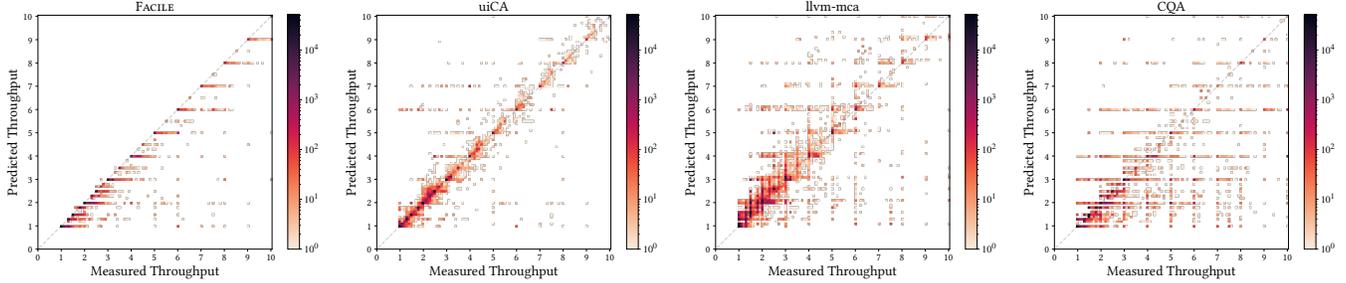

\centering
\begin{subfigure}[t]{0.24\textwidth}\resizebox{\textwidth}{!}{\import{figures/heatmaps}{hm_rkl_loop_Analytical.pgf}}\end{subfigure}~
\begin{subfigure}[t]{0.24\textwidth}\resizebox{\textwidth}{!}{\import{figures/heatmaps}{hm_rkl_loop_uiCA.pgf}}\end{subfigure}~
\begin{subfigure}[t]{0.24\textwidth}\resizebox{\textwidth}{!}{\import{figures/heatmaps}{hm_rkl_loop_llvm-mca.pgf}}\end{subfigure}~
\begin{subfigure}[t]{0.24\textwidth}\resizebox{\textwidth}{!}{\import{figures/heatmaps}{hm_rkl_loop_CQA.pgf}}\end{subfigure}
\caption{Heatmaps for \bhivel for basic blocks with a measured throughput of less than 10 cycles/iteration on Rocket Lake}
\label{fig:heatmaps}
\end{figure*}

Figure~\ref{fig:heatmaps} shows heatmaps relating the measured and predicted throughputs for \bhivel benchmarks with a measured throughput of less than 10 cycles on Rocket Lake.

The key takeaway from Figure~\ref{fig:heatmaps} is that both \toolname and uiCA are able to predict the throughput of most benchmarks with a high accuracy.
We also note that \toolname is always optimistic in its predictions, i.e., it always predicts a higher throughput than the measured throughput.


\newcommand{\STAB}[1]{\begin{tabular}{@{}c@{}}#1\end{tabular}}

\begin{table}
\caption{Influence of components on the prediction accuracy for Rocket Lake, Skylake, and Sandy Bridge.}
\negvspace
\label{tab:analytical-components}
\begin{center}
\begin{tabular}{llrcrc}
\toprule
& & \multicolumn{2}{c}{\textbf{\bhiveu}} & \multicolumn{2}{c}{\textbf{\bhivel}}\\  \cmidrule(lr){3-4}\cmidrule(lr){5-6}
 & \textbf{Predictor} & \hspace{-2mm}\textbf{MAPE} & \hspace{-4mm}\textbf{Kendall}\hspace{-2mm} & \hspace{-2mm}\textbf{MAPE} & \hspace{-4mm}\textbf{Kendall}\hspace{-2mm} \\
\midrule
\hspace{-1mm}\multirow{19}{*}{\STAB{\rotatebox[origin=c]{90}{RKL}}}\hspace{-3mm} & \toolname & 0.42\% & 0.9860 & 1.04\% & 0.9731\\
 & \toolname w/ \simplepredec\hspace{-8mm}  & 4.38\% & 0.9024 & & \\
 & \toolname w/ \simpledec  & 0.69\% & 0.9836 & & \\
 & only \predec  & 14.84\% & 0.7291 & & \\
 & only \dec  & 26.37\% & 0.6840 & & \\
 & only \dsb  & & & 100.00\% & 0.0078\\
 & only \lsd  & & & 24.26\% & 0.6505\\
 & only \issue  & 41.45\% & 0.6774 & 29.85\% & 0.6520\\
 & only \pu  & 29.18\% & 0.5122 & 26.40\% & 0.7132\\
 & only \lat  & 88.56\% & 0.3589 & 22.55\% & 0.4428\\
 & only {\predec}+\pu\hspace{-5mm}  & 9.36\% & 0.8399 & & \\
 & only {\lat}+\pu  & 22.40\% & 0.6149 & 6.59\% & 0.8467\\
 & \toolname w/o \predec  & 9.02\% & 0.8512 & & \\
 & \toolname w/o \dec  & 1.19\% & 0.9734 & & \\
 & \toolname w/o \dsb  & & & 1.04\% & 0.9731\\
 & \toolname w/o \lsd  & & & 2.52\% & 0.9556\\
 & \toolname w/o \issue  & 0.43\% & 0.9857 & 1.06\% & 0.9724\\
 & \toolname w/o \pu  & 7.88\% & 0.8570 & 7.90\% & 0.8002\\
 & \toolname w/o \lat  & 5.36\% & 0.8853 & 14.88\% & 0.7967\\
 \midrule
 \hspace{-1mm}\multirow{3}{*}{\STAB{\rotatebox[origin=c]{90}{SKL}}}\hspace{-3mm} & \toolname & 0.69\% & 0.9753 & 1.08\% & 0.9752\\
 & \toolname w/o \dsb  & & & 2.70\% & 0.9629\\
 & \toolname w/o \issue  & 0.76\% & 0.9736 & 4.34\% & 0.9107\\
 \midrule
 \hspace{-1mm}\multirow{2}{*}{\STAB{\rotatebox[origin=c]{90}{SNB}}}\hspace{-3mm} & \toolname & 1.95\% & 0.9586 & 1.33\% & 0.9742\\
 & \toolname w/ \simpledec  & 3.76\% & 0.9378 & & \\
 \bottomrule
\end{tabular}
\end{center}
\vspace{-5mm}
\end{table}

\subsubsection*{Influence of \toolname's components on accuracy}

Next, we study the importance of each component of \toolname for the purpose of accurate throughput prediction.
Table~\ref{tab:analytical-components} shows the MAPE and Kendall's tau for the different variants of \toolname on both \bhiveu and \bhivel benchmarks for Rocket Lake, Skylake, and Sandy Bridge.
Note that for components that are not used by \DefTPU or \DefTPL, the corresponding cells in the table are empty.

First, consider the experiments where we replace individual components of \toolname with simpler versions.
Under Rocket Lake, replacing \predec with \simplepredec leads to a significant drop in accuracy, which shows that the more complex \predec is required for accurate throughput prediction.
Replacing \dec by \simpledec has a smaller impact on accuracy for Rocket Lake.
However, note the significant drop in accuracy for Sandy Bridge.

Next, consider the experiments where we analyze the accuracy of individual components as standalone predictors (rows ``only X'').
We can see that neither of the components alone can predict throughput accurately.
Combining the two most accurate components, {\predec}+\pu for \bhiveu and {\lat}+\pu for \bhivel, yields significantly higher accuracy but still clearly falls short of \toolname as a whole.

Let us now turn to the experiments where we exclude individual components from \toolname (rows ``\dots w/o X'').
On Rocket Lake, we see a notable drop in accuracy when excluding \predec, \pu, and \lat under \bhiveu and/or \bhivel.
Excluding \dec and \lsd has a smaller, yet non-negligible impact on accuracy.
In contrast, excluding \issue and \dsb has almost no effect on accuracy.
For Rocket Lake, individually, these components are thus not required for accurate throughput prediction. 
However, for Skylake, we see that excluding \issue or \dsb leads to a significant drop in accuracy under \bhivel, highlighting their importance in general.\looseness=-1

\subsection{Computational Efficiency}

In this section, we study the computational efficiency of \toolname to explore its suitability for use cases that require fast throughput prediction.
We measure and compare the execution time of \toolname and the other predictors on both the \bhiveu and \bhivel benchmarks.

We conduct our experiments on a Linux workstation with an AMD Ryzen 5900X CPU running at \SI{3.70}{\giga\hertz} with \SI{64}{\giga\byte} of RAM.
To ensure consistency of measured execution times, we disable simultaneous multithreading (SMT), set the scaling governor to `performance', and disable dynamic frequency scaling due to Turbo Boost.
We run each predictor sequentially on all benchmarks for the Skylake microarchitecture and measure the execution time.

\begin{figure*}[ht]
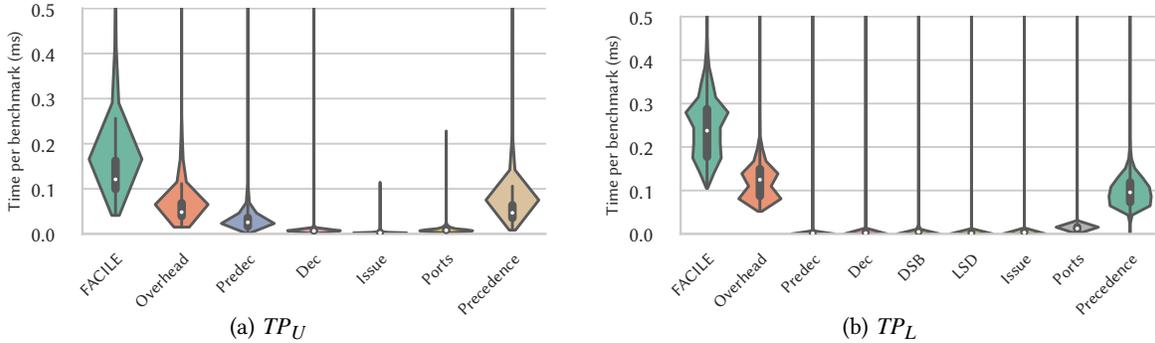

  \hfill
  \begin{subfigure}[t]{0.45\linewidth}
    \centering
  \resizebox{\linewidth}{!}{\input{figures/time_unroll_dist.pgf}}
  \vspace*{-12mm}
  \caption{\DefTPU}
  \label{fig:componentwise-breakdown-tpu}
  \end{subfigure}
  \begin{subfigure}[t]{0.45\linewidth}
    \centering
  \resizebox{\linewidth}{!}{\input{figures/time_loop_dist.pgf}}
  \vspace*{-12mm}
  \caption{\DefTPL}
  \label{fig:componentwise-breakdown-tpl}
  \end{subfigure}
  \hfill~
  \caption{Distributions of execution times of \toolname's components under \DefTPU and \DefTPL.}
  \label{fig:componentwise-breakdown}
\end{figure*}

\begin{figure}[h]
  \centering
  \resizebox{\linewidth}{!}{\input{figures/tools_comparison.pgf}}
  \caption{Efficiency of \toolname compared to other tools.}
  \label{fig:runtime-compared}
\end{figure}
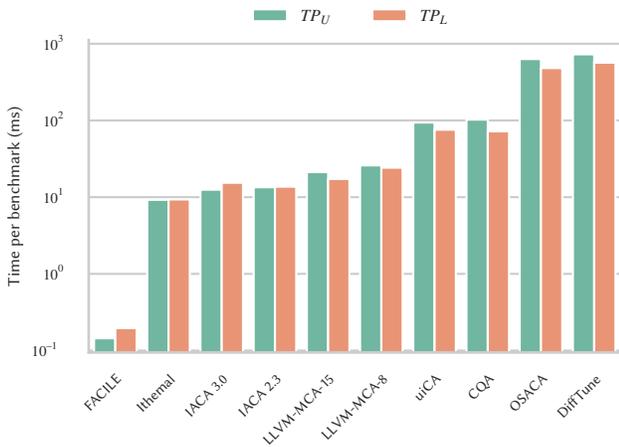

\subsubsection*{\toolname versus other predictors}
Figure~\ref{fig:runtime-compared} shows the execution time of the different predictors on the \bhiveu and \bhivel benchmarks.
We observe that \toolname is faster than all existing predictors by orders of magnitude.
The closest competitor is Ithemal, which is still almost two orders of magnitude slower than \toolname.

Python-based predictors like OSACA, uiCA, Ithemal and \toolname incur an overhead every time they are run due to the boot up time of the Python VM.
All the other predictors in our experiments are written in C/C++ and do not have this overhead.
Both OSACA and uiCA are run once for each benchmark, hence the overhead is incurred for each benchmark.
For \toolname, this overhead is only incurred once as we pass all benchmarks to the predictor in a single run.
We run Ithemal in interactive mode, using scripts from~\cite{uica-eval} that allow us to load the predictor once and keep passing benchmarks to it.
Hence, the Python VM bootup time is incurred only once for Ithemal also.
However, even if we deduct the boot up time from OSACA and uiCA’s runtime, they are still slower than \toolname by more than a factor of 100.\looseness=-1 
\todo{we learn here in passing that \toolname is implemented in Python. there should probably be a small paragraph about the implementation somewhere. what else is there to say? parsing of inputs, use of uops.info data, etc.? does it make sense to have a block diagram with the most important components?}

\subsubsection*{Influence of \toolname's components on efficiency}

\begin{figure*}[h]
  \hfill
  \begin{subfigure}[t]{0.6725\columnwidth}
    \includegraphics[width=\linewidth]{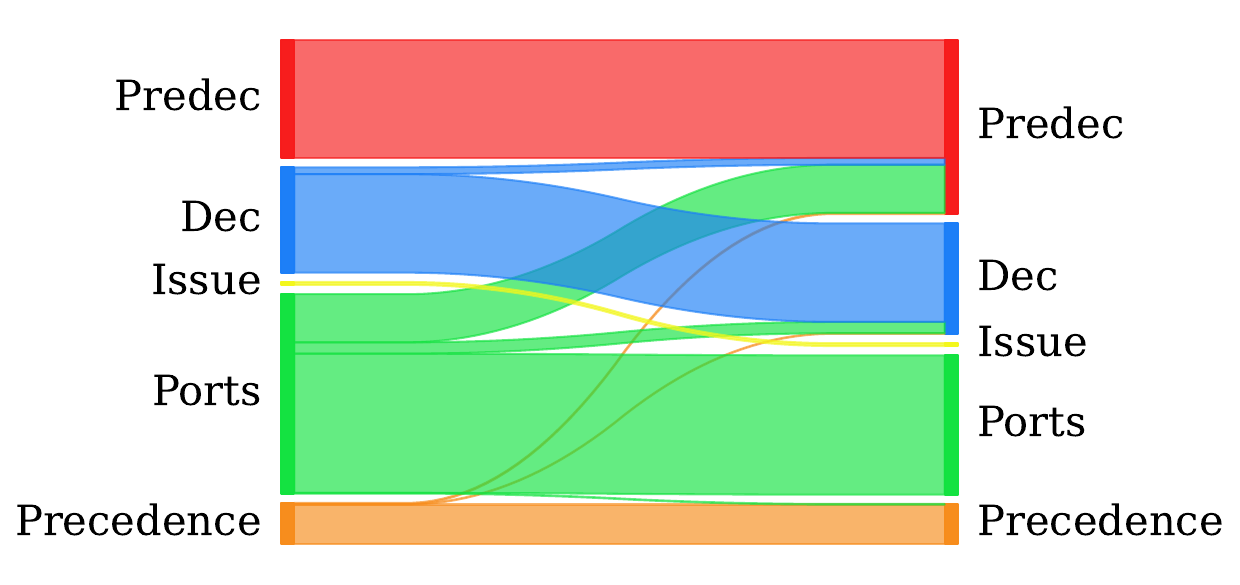}
    \caption{From Sandy Bridge to Haswell}
  \end{subfigure}
  \begin{subfigure}[t]{0.6725\columnwidth}
    \includegraphics[width=\linewidth]{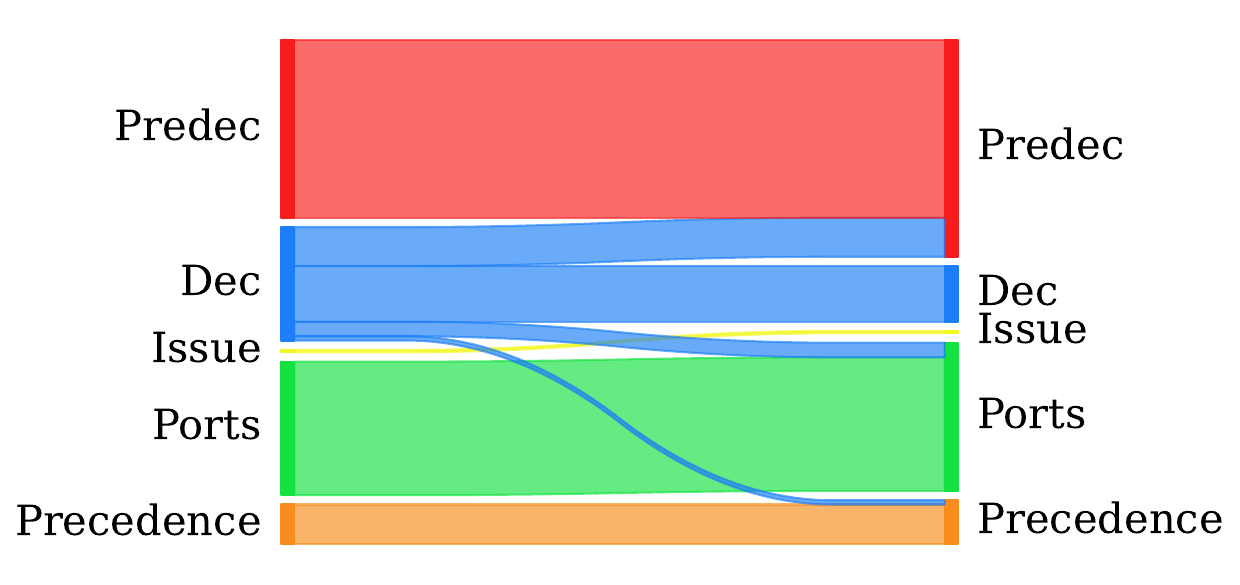}
    \caption{From Haswell to Cascade Lake}
  \end{subfigure}
  \begin{subfigure}[t]{0.6725\columnwidth}
    \includegraphics[width=\linewidth]{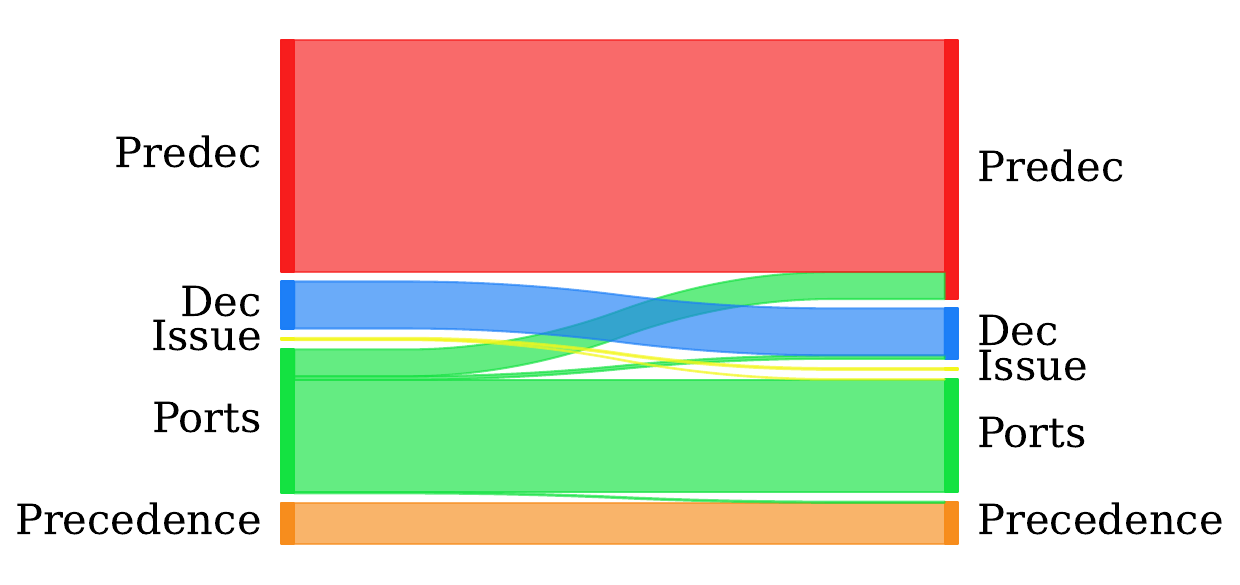}
    \caption{From Cascade Lake to Rocket Lake}
  \end{subfigure}
  \hfill~
\caption{Evolution of bottlenecks under \DefTPU from Sandy Bridge to Rocket Lake via  Haswell and Cascade Lake.}\label{fig:bottleneck-evolution}
\end{figure*}

We also measure the time taken by \toolname to compute the individual components.
First, we run \toolname on all benchmarks with all components deactivated to get the time spent on overhead like parsing the input, disassembling the benchmark, etc.
Then, we run individual components on all benchmarks and deduct the overhead runtime to get the time taken to compute each component.


Figure~\ref{fig:componentwise-breakdown} shows the distributions of execution times of \toolname's components under \DefTPU and \DefTPL. 
The overhead and the \lat component take up close to 90\% of the total execution time in both cases.
This indicates that any further efforts to improve the efficiency of \toolname will first need to be focussed on these two factors.
\predec and \dec take less time for \DefTPL than for \DefTPU as they are often skipped for the loop benchmarks (see Equation~\ref{eq:fe}).
The loop benchmarks are generally larger than the unrolled benchmarks, which explains why their analysis takes longer on the average.\todo{it's a bit strange that \toolname is faster on \DefTPU in contrast to almost all other tools.}

\subsection{Exploiting Interpretability}

In this section, we exploit the interpretability of \toolname to gain insights into the evolution of Intel microarchitectures and the impact of individual components on the overall performance.
For brevity, we limit this analysis to \DefTPU.\looseness=-1

\subsubsection*{Evolution of bottlenecks across microarchitectures}

First, we analyze the evolution of bottlenecks across microarchitectures over the last decade on \bhiveu from Sandy Bridge via Haswell and Cascade Lake to Rocket Lake.
For each \bhiveu benchmark, we determine the bottleneck component for each of the considered microarchitectures.
In case multiple components induce the same throughput, we consider the component that is closest to the front end as the bottleneck, i.e., $\predec > \dec > \issue > \pu > \lat$.
The Sankey diagram in Figure~\ref{fig:bottleneck-evolution} shows the share of benchmarks for which a particular component was the bottleneck and how this share evolved across microarchitectures.
We can observe that the share of benchmarks that are \predec-bound has increased over time, while the share of benchmarks that are \pu-bound has decreased.\looseness=-1

\subsubsection*{Potential for performance improvement}

\begin{table}
  \centering
  \caption{Speedup when idealizing a single component.}\label{tab:component-impact}
\begin{tabular}{lccccc}
  \toprule
   &  \predec &  \dec &  \issue &  \pu &  \lat \\
  \midrule
  SNB &  1.04 &  1.03 &  1.00 &  1.17 &  1.08 \\
  IVB &  1.07 &  1.02 &  1.00 &  1.16 &  1.09 \\
  HSW &  1.07 &  1.03 &  1.00 &  1.13 &  1.08 \\
  BDW &  1.08 &  1.03 &  1.00 &  1.12 &  1.08 \\
  SKL &  1.11 &  1.00 &  1.00 &  1.17 &  1.11 \\
  CLX &  1.10 &  1.00 &  1.00 &  1.15 &  1.09 \\
  ICL &  1.11 &  1.01 &  1.00 &  1.08 &  1.12 \\
  TGL &  1.11 &  1.01 &  1.00 &  1.08 &  1.12 \\
  RKL &  1.12 &  1.01 &  1.00 &  1.10 &  1.10 \\
  \bottomrule
\end{tabular}
\end{table}

Finally, we use the model to answer the following counterfactual question:
How would performance improve if a particular component were made infinitely fast?
Table~\ref{tab:component-impact} answers this question for all microarchitectures from Sandy Bridge to Rocket Lake.
As expected from the previous experiment, the potential for performance improvement has gradually shifted towards \predec over time.
However, as the microarchitectural designs are balanced, performance is usually similarly constrained by multiple components.
As a consequence, the potential for performance gains by improving a single component only is limited.




\section{Conclusions and Future Work}

Our experiments demonstrate that \toolname's compositional analytical model enables highly-accurate, efficient, and interpretable basic-block throughput prediction.

\toolname's efficiency makes it particularly suitable for use cases that require fast throughput prediction, e.g., in superoptimizers.  
At first glance, interpretability may seem more relevant for human users than for such automatic tools. 
However, we believe that interpretability may also be beneficial in settings with no humans in the loop.
For example, a superoptimizer could use the model to identify bottlenecks and prioritize optimizations accordingly. 
Exploring this idea is a direction for future work.\looseness=-1

Another direction for future work is to extend \toolname to handle more complex code, e.g., involving branches, and to lift some of the modeling assumptions, e.g., regarding the absence of cache misses.
To this end, it may be profitable to combine \toolname's static predictions with information gathered from dynamic profiling.



\todo{what are the limitations of our approach? can they be overcome? potential next steps?}
\todo{what are the things that are modeled in uiCA, but not in \toolname? why are they irrelevant? when might they be relevant? one thing that comes to mind are buffers (and their sizes)}


\todo{mention superoptimizers again; ``interpretability is not just for humans; it could also help superoptimizers or other automatic tools''}

\todo{what about branches and cache misses? more obvious to handle in simulator!?}

\section*{Acknowledgments}
This project has received funding from the European Research Council under the Horizon 2020 research and innovation programme (grant agreement No. 101020415).

\bibliographystyle{IEEEtranS}
\bibliography{references}

\end{document}

%% file: figures/tools_comparison.pgf
\begingroup%
\makeatletter%
\begin{pgfpicture}%
\pgfpathrectangle{\pgfpointorigin}{\pgfqpoint{4.204300in}{3.153974in}}%
\pgfusepath{use as bounding box, clip}%
\begin{pgfscope}%
\pgfsetbuttcap%
\pgfsetmiterjoin%
\definecolor{currentfill}{rgb}{1.000000,1.000000,1.000000}%
\pgfsetfillcolor{currentfill}%
\pgfsetlinewidth{0.000000pt}%
\definecolor{currentstroke}{rgb}{1.000000,1.000000,1.000000}%
\pgfsetstrokecolor{currentstroke}%
\pgfsetdash{}{0pt}%
\pgfpathmoveto{\pgfqpoint{0.000000in}{0.000000in}}%
\pgfpathlineto{\pgfqpoint{4.204300in}{0.000000in}}%
\pgfpathlineto{\pgfqpoint{4.204300in}{3.153974in}}%
\pgfpathlineto{\pgfqpoint{0.000000in}{3.153974in}}%
\pgfpathlineto{\pgfqpoint{0.000000in}{0.000000in}}%
\pgfpathclose%
\pgfusepath{fill}%
\end{pgfscope}%
\begin{pgfscope}%
\pgfsetbuttcap%
\pgfsetmiterjoin%
\definecolor{currentfill}{rgb}{1.000000,1.000000,1.000000}%
\pgfsetfillcolor{currentfill}%
\pgfsetlinewidth{0.000000pt}%
\definecolor{currentstroke}{rgb}{0.000000,0.000000,0.000000}%
\pgfsetstrokecolor{currentstroke}%
\pgfsetstrokeopacity{0.000000}%
\pgfsetdash{}{0pt}%
\pgfpathmoveto{\pgfqpoint{0.635194in}{0.763790in}}%
\pgfpathlineto{\pgfqpoint{4.104300in}{0.763790in}}%
\pgfpathlineto{\pgfqpoint{4.104300in}{2.802000in}}%
\pgfpathlineto{\pgfqpoint{0.635194in}{2.802000in}}%
\pgfpathlineto{\pgfqpoint{0.635194in}{0.763790in}}%
\pgfpathclose%
\pgfusepath{fill}%
\end{pgfscope}%
\begin{pgfscope}%
\definecolor{textcolor}{rgb}{0.150000,0.150000,0.150000}%
\pgfsetstrokecolor{textcolor}%
\pgfsetfillcolor{textcolor}%
\pgftext[x=0.808650in,y=0.631845in,right,top,rotate=45.000000]{\color{textcolor}\sffamily\fontsize{7.000000}{8.400000}\selectfont FACILE}%
\end{pgfscope}%
\begin{pgfscope}%
\definecolor{textcolor}{rgb}{0.150000,0.150000,0.150000}%
\pgfsetstrokecolor{textcolor}%
\pgfsetfillcolor{textcolor}%
\pgftext[x=1.155560in,y=0.631845in,right,top,rotate=45.000000]{\color{textcolor}\sffamily\fontsize{7.000000}{8.400000}\selectfont Ithemal}%
\end{pgfscope}%
\begin{pgfscope}%
\definecolor{textcolor}{rgb}{0.150000,0.150000,0.150000}%
\pgfsetstrokecolor{textcolor}%
\pgfsetfillcolor{textcolor}%
\pgftext[x=1.502471in,y=0.631845in,right,top,rotate=45.000000]{\color{textcolor}\sffamily\fontsize{7.000000}{8.400000}\selectfont IACA 3.0}%
\end{pgfscope}%
\begin{pgfscope}%
\definecolor{textcolor}{rgb}{0.150000,0.150000,0.150000}%
\pgfsetstrokecolor{textcolor}%
\pgfsetfillcolor{textcolor}%
\pgftext[x=1.849381in,y=0.631845in,right,top,rotate=45.000000]{\color{textcolor}\sffamily\fontsize{7.000000}{8.400000}\selectfont IACA 2.3}%
\end{pgfscope}%
\begin{pgfscope}%
\definecolor{textcolor}{rgb}{0.150000,0.150000,0.150000}%
\pgfsetstrokecolor{textcolor}%
\pgfsetfillcolor{textcolor}%
\pgftext[x=2.196292in,y=0.631845in,right,top,rotate=45.000000]{\color{textcolor}\sffamily\fontsize{7.000000}{8.400000}\selectfont LLVM-MCA-15}%
\end{pgfscope}%
\begin{pgfscope}%
\definecolor{textcolor}{rgb}{0.150000,0.150000,0.150000}%
\pgfsetstrokecolor{textcolor}%
\pgfsetfillcolor{textcolor}%
\pgftext[x=2.543202in,y=0.631845in,right,top,rotate=45.000000]{\color{textcolor}\sffamily\fontsize{7.000000}{8.400000}\selectfont LLVM-MCA-8}%
\end{pgfscope}%
\begin{pgfscope}%
\definecolor{textcolor}{rgb}{0.150000,0.150000,0.150000}%
\pgfsetstrokecolor{textcolor}%
\pgfsetfillcolor{textcolor}%
\pgftext[x=2.890113in,y=0.631845in,right,top,rotate=45.000000]{\color{textcolor}\sffamily\fontsize{7.000000}{8.400000}\selectfont uiCA}%
\end{pgfscope}%
\begin{pgfscope}%
\definecolor{textcolor}{rgb}{0.150000,0.150000,0.150000}%
\pgfsetstrokecolor{textcolor}%
\pgfsetfillcolor{textcolor}%
\pgftext[x=3.237023in,y=0.631845in,right,top,rotate=45.000000]{\color{textcolor}\sffamily\fontsize{7.000000}{8.400000}\selectfont CQA}%
\end{pgfscope}%
\begin{pgfscope}%
\definecolor{textcolor}{rgb}{0.150000,0.150000,0.150000}%
\pgfsetstrokecolor{textcolor}%
\pgfsetfillcolor{textcolor}%
\pgftext[x=3.583934in,y=0.631845in,right,top,rotate=45.000000]{\color{textcolor}\sffamily\fontsize{7.000000}{8.400000}\selectfont OSACA}%
\end{pgfscope}%
\begin{pgfscope}%
\definecolor{textcolor}{rgb}{0.150000,0.150000,0.150000}%
\pgfsetstrokecolor{textcolor}%
\pgfsetfillcolor{textcolor}%
\pgftext[x=3.930844in,y=0.631845in,right,top,rotate=45.000000]{\color{textcolor}\sffamily\fontsize{7.000000}{8.400000}\selectfont DiffTune}%
\end{pgfscope}%
\begin{pgfscope}%
\pgfpathrectangle{\pgfqpoint{0.635194in}{0.763790in}}{\pgfqpoint{3.469105in}{2.038210in}}%
\pgfusepath{clip}%
\pgfsetroundcap%
\pgfsetroundjoin%
\pgfsetlinewidth{1.003750pt}%
\definecolor{currentstroke}{rgb}{0.800000,0.800000,0.800000}%
\pgfsetstrokecolor{currentstroke}%
\pgfsetdash{}{0pt}%
\pgfpathmoveto{\pgfqpoint{0.635194in}{0.776313in}}%
\pgfpathlineto{\pgfqpoint{4.104300in}{0.776313in}}%
\pgfusepath{stroke}%
\end{pgfscope}%
\begin{pgfscope}%
\definecolor{textcolor}{rgb}{0.150000,0.150000,0.150000}%
\pgfsetstrokecolor{textcolor}%
\pgfsetfillcolor{textcolor}%
\pgftext[x=0.263086in, y=0.738847in, left, base]{\color{textcolor}\sffamily\fontsize{7.000000}{8.400000}\selectfont \(\displaystyle {10^{-1}}\)}%
\end{pgfscope}%
\begin{pgfscope}%
\pgfpathrectangle{\pgfqpoint{0.635194in}{0.763790in}}{\pgfqpoint{3.469105in}{2.038210in}}%
\pgfusepath{clip}%
\pgfsetroundcap%
\pgfsetroundjoin%
\pgfsetlinewidth{1.003750pt}%
\definecolor{currentstroke}{rgb}{0.800000,0.800000,0.800000}%
\pgfsetstrokecolor{currentstroke}%
\pgfsetdash{}{0pt}%
\pgfpathmoveto{\pgfqpoint{0.635194in}{1.276800in}}%
\pgfpathlineto{\pgfqpoint{4.104300in}{1.276800in}}%
\pgfusepath{stroke}%
\end{pgfscope}%
\begin{pgfscope}%
\definecolor{textcolor}{rgb}{0.150000,0.150000,0.150000}%
\pgfsetstrokecolor{textcolor}%
\pgfsetfillcolor{textcolor}%
\pgftext[x=0.338318in, y=1.239867in, left, base]{\color{textcolor}\sffamily\fontsize{7.000000}{8.400000}\selectfont \(\displaystyle {10^{0}}\)}%
\end{pgfscope}%
\begin{pgfscope}%
\pgfpathrectangle{\pgfqpoint{0.635194in}{0.763790in}}{\pgfqpoint{3.469105in}{2.038210in}}%
\pgfusepath{clip}%
\pgfsetroundcap%
\pgfsetroundjoin%
\pgfsetlinewidth{1.003750pt}%
\definecolor{currentstroke}{rgb}{0.800000,0.800000,0.800000}%
\pgfsetstrokecolor{currentstroke}%
\pgfsetdash{}{0pt}%
\pgfpathmoveto{\pgfqpoint{0.635194in}{1.777287in}}%
\pgfpathlineto{\pgfqpoint{4.104300in}{1.777287in}}%
\pgfusepath{stroke}%
\end{pgfscope}%
\begin{pgfscope}%
\definecolor{textcolor}{rgb}{0.150000,0.150000,0.150000}%
\pgfsetstrokecolor{textcolor}%
\pgfsetfillcolor{textcolor}%
\pgftext[x=0.338318in, y=1.740354in, left, base]{\color{textcolor}\sffamily\fontsize{7.000000}{8.400000}\selectfont \(\displaystyle {10^{1}}\)}%
\end{pgfscope}%
\begin{pgfscope}%
\pgfpathrectangle{\pgfqpoint{0.635194in}{0.763790in}}{\pgfqpoint{3.469105in}{2.038210in}}%
\pgfusepath{clip}%
\pgfsetroundcap%
\pgfsetroundjoin%
\pgfsetlinewidth{1.003750pt}%
\definecolor{currentstroke}{rgb}{0.800000,0.800000,0.800000}%
\pgfsetstrokecolor{currentstroke}%
\pgfsetdash{}{0pt}%
\pgfpathmoveto{\pgfqpoint{0.635194in}{2.277774in}}%
\pgfpathlineto{\pgfqpoint{4.104300in}{2.277774in}}%
\pgfusepath{stroke}%
\end{pgfscope}%
\begin{pgfscope}%
\definecolor{textcolor}{rgb}{0.150000,0.150000,0.150000}%
\pgfsetstrokecolor{textcolor}%
\pgfsetfillcolor{textcolor}%
\pgftext[x=0.338318in, y=2.240841in, left, base]{\color{textcolor}\sffamily\fontsize{7.000000}{8.400000}\selectfont \(\displaystyle {10^{2}}\)}%
\end{pgfscope}%
\begin{pgfscope}%
\pgfpathrectangle{\pgfqpoint{0.635194in}{0.763790in}}{\pgfqpoint{3.469105in}{2.038210in}}%
\pgfusepath{clip}%
\pgfsetroundcap%
\pgfsetroundjoin%
\pgfsetlinewidth{1.003750pt}%
\definecolor{currentstroke}{rgb}{0.800000,0.800000,0.800000}%
\pgfsetstrokecolor{currentstroke}%
\pgfsetdash{}{0pt}%
\pgfpathmoveto{\pgfqpoint{0.635194in}{2.778261in}}%
\pgfpathlineto{\pgfqpoint{4.104300in}{2.778261in}}%
\pgfusepath{stroke}%
\end{pgfscope}%
\begin{pgfscope}%
\definecolor{textcolor}{rgb}{0.150000,0.150000,0.150000}%
\pgfsetstrokecolor{textcolor}%
\pgfsetfillcolor{textcolor}%
\pgftext[x=0.338318in, y=2.741328in, left, base]{\color{textcolor}\sffamily\fontsize{7.000000}{8.400000}\selectfont \(\displaystyle {10^{3}}\)}%
\end{pgfscope}%
\begin{pgfscope}%
\definecolor{textcolor}{rgb}{0.150000,0.150000,0.150000}%
\pgfsetstrokecolor{textcolor}%
\pgfsetfillcolor{textcolor}%
\pgftext[x=0.207530in,y=1.782895in,,bottom,rotate=90.000000]{\color{textcolor}\sffamily\fontsize{8.000000}{9.600000}\selectfont Time per benchmark (ms)}%
\end{pgfscope}%
\begin{pgfscope}%
\pgfpathrectangle{\pgfqpoint{0.635194in}{0.763790in}}{\pgfqpoint{3.469105in}{2.038210in}}%
\pgfusepath{clip}%
\pgfsetbuttcap%
\pgfsetmiterjoin%
\definecolor{currentfill}{rgb}{0.445098,0.715686,0.630392}%
\pgfsetfillcolor{currentfill}%
\pgfsetlinewidth{1.003750pt}%
\definecolor{currentstroke}{rgb}{1.000000,1.000000,1.000000}%
\pgfsetstrokecolor{currentstroke}%
\pgfsetdash{}{0pt}%
\pgfpathmoveto{\pgfqpoint{0.808650in}{-226.701573in}}%
\pgfpathlineto{\pgfqpoint{0.808650in}{0.856436in}}%
\pgfpathlineto{\pgfqpoint{0.669885in}{0.856436in}}%
\pgfpathlineto{\pgfqpoint{0.669885in}{-226.701573in}}%
\pgfusepath{stroke,fill}%
\end{pgfscope}%
\begin{pgfscope}%
\pgfpathrectangle{\pgfqpoint{0.635194in}{0.763790in}}{\pgfqpoint{3.469105in}{2.038210in}}%
\pgfusepath{clip}%
\pgfsetbuttcap%
\pgfsetmiterjoin%
\definecolor{currentfill}{rgb}{0.445098,0.715686,0.630392}%
\pgfsetfillcolor{currentfill}%
\pgfsetlinewidth{1.003750pt}%
\definecolor{currentstroke}{rgb}{1.000000,1.000000,1.000000}%
\pgfsetstrokecolor{currentstroke}%
\pgfsetdash{}{0pt}%
\pgfpathmoveto{\pgfqpoint{1.155560in}{-226.701573in}}%
\pgfpathlineto{\pgfqpoint{1.155560in}{1.759929in}}%
\pgfpathlineto{\pgfqpoint{1.016796in}{1.759929in}}%
\pgfpathlineto{\pgfqpoint{1.016796in}{-226.701573in}}%
\pgfusepath{stroke,fill}%
\end{pgfscope}%
\begin{pgfscope}%
\pgfpathrectangle{\pgfqpoint{0.635194in}{0.763790in}}{\pgfqpoint{3.469105in}{2.038210in}}%
\pgfusepath{clip}%
\pgfsetbuttcap%
\pgfsetmiterjoin%
\definecolor{currentfill}{rgb}{0.445098,0.715686,0.630392}%
\pgfsetfillcolor{currentfill}%
\pgfsetlinewidth{1.003750pt}%
\definecolor{currentstroke}{rgb}{1.000000,1.000000,1.000000}%
\pgfsetstrokecolor{currentstroke}%
\pgfsetdash{}{0pt}%
\pgfpathmoveto{\pgfqpoint{1.502471in}{-226.701573in}}%
\pgfpathlineto{\pgfqpoint{1.502471in}{1.826140in}}%
\pgfpathlineto{\pgfqpoint{1.363707in}{1.826140in}}%
\pgfpathlineto{\pgfqpoint{1.363707in}{-226.701573in}}%
\pgfusepath{stroke,fill}%
\end{pgfscope}%
\begin{pgfscope}%
\pgfpathrectangle{\pgfqpoint{0.635194in}{0.763790in}}{\pgfqpoint{3.469105in}{2.038210in}}%
\pgfusepath{clip}%
\pgfsetbuttcap%
\pgfsetmiterjoin%
\definecolor{currentfill}{rgb}{0.445098,0.715686,0.630392}%
\pgfsetfillcolor{currentfill}%
\pgfsetlinewidth{1.003750pt}%
\definecolor{currentstroke}{rgb}{1.000000,1.000000,1.000000}%
\pgfsetstrokecolor{currentstroke}%
\pgfsetdash{}{0pt}%
\pgfpathmoveto{\pgfqpoint{1.849381in}{-226.701573in}}%
\pgfpathlineto{\pgfqpoint{1.849381in}{1.841920in}}%
\pgfpathlineto{\pgfqpoint{1.710617in}{1.841920in}}%
\pgfpathlineto{\pgfqpoint{1.710617in}{-226.701573in}}%
\pgfusepath{stroke,fill}%
\end{pgfscope}%
\begin{pgfscope}%
\pgfpathrectangle{\pgfqpoint{0.635194in}{0.763790in}}{\pgfqpoint{3.469105in}{2.038210in}}%
\pgfusepath{clip}%
\pgfsetbuttcap%
\pgfsetmiterjoin%
\definecolor{currentfill}{rgb}{0.445098,0.715686,0.630392}%
\pgfsetfillcolor{currentfill}%
\pgfsetlinewidth{1.003750pt}%
\definecolor{currentstroke}{rgb}{1.000000,1.000000,1.000000}%
\pgfsetstrokecolor{currentstroke}%
\pgfsetdash{}{0pt}%
\pgfpathmoveto{\pgfqpoint{2.196292in}{-226.701573in}}%
\pgfpathlineto{\pgfqpoint{2.196292in}{1.940548in}}%
\pgfpathlineto{\pgfqpoint{2.057528in}{1.940548in}}%
\pgfpathlineto{\pgfqpoint{2.057528in}{-226.701573in}}%
\pgfusepath{stroke,fill}%
\end{pgfscope}%
\begin{pgfscope}%
\pgfpathrectangle{\pgfqpoint{0.635194in}{0.763790in}}{\pgfqpoint{3.469105in}{2.038210in}}%
\pgfusepath{clip}%
\pgfsetbuttcap%
\pgfsetmiterjoin%
\definecolor{currentfill}{rgb}{0.445098,0.715686,0.630392}%
\pgfsetfillcolor{currentfill}%
\pgfsetlinewidth{1.003750pt}%
\definecolor{currentstroke}{rgb}{1.000000,1.000000,1.000000}%
\pgfsetstrokecolor{currentstroke}%
\pgfsetdash{}{0pt}%
\pgfpathmoveto{\pgfqpoint{2.543202in}{-226.701573in}}%
\pgfpathlineto{\pgfqpoint{2.543202in}{1.984538in}}%
\pgfpathlineto{\pgfqpoint{2.404438in}{1.984538in}}%
\pgfpathlineto{\pgfqpoint{2.404438in}{-226.701573in}}%
\pgfusepath{stroke,fill}%
\end{pgfscope}%
\begin{pgfscope}%
\pgfpathrectangle{\pgfqpoint{0.635194in}{0.763790in}}{\pgfqpoint{3.469105in}{2.038210in}}%
\pgfusepath{clip}%
\pgfsetbuttcap%
\pgfsetmiterjoin%
\definecolor{currentfill}{rgb}{0.445098,0.715686,0.630392}%
\pgfsetfillcolor{currentfill}%
\pgfsetlinewidth{1.003750pt}%
\definecolor{currentstroke}{rgb}{1.000000,1.000000,1.000000}%
\pgfsetstrokecolor{currentstroke}%
\pgfsetdash{}{0pt}%
\pgfpathmoveto{\pgfqpoint{2.890113in}{-226.701573in}}%
\pgfpathlineto{\pgfqpoint{2.890113in}{2.264137in}}%
\pgfpathlineto{\pgfqpoint{2.751349in}{2.264137in}}%
\pgfpathlineto{\pgfqpoint{2.751349in}{-226.701573in}}%
\pgfusepath{stroke,fill}%
\end{pgfscope}%
\begin{pgfscope}%
\pgfpathrectangle{\pgfqpoint{0.635194in}{0.763790in}}{\pgfqpoint{3.469105in}{2.038210in}}%
\pgfusepath{clip}%
\pgfsetbuttcap%
\pgfsetmiterjoin%
\definecolor{currentfill}{rgb}{0.445098,0.715686,0.630392}%
\pgfsetfillcolor{currentfill}%
\pgfsetlinewidth{1.003750pt}%
\definecolor{currentstroke}{rgb}{1.000000,1.000000,1.000000}%
\pgfsetstrokecolor{currentstroke}%
\pgfsetdash{}{0pt}%
\pgfpathmoveto{\pgfqpoint{3.237023in}{-226.701573in}}%
\pgfpathlineto{\pgfqpoint{3.237023in}{2.283302in}}%
\pgfpathlineto{\pgfqpoint{3.098259in}{2.283302in}}%
\pgfpathlineto{\pgfqpoint{3.098259in}{-226.701573in}}%
\pgfusepath{stroke,fill}%
\end{pgfscope}%
\begin{pgfscope}%
\pgfpathrectangle{\pgfqpoint{0.635194in}{0.763790in}}{\pgfqpoint{3.469105in}{2.038210in}}%
\pgfusepath{clip}%
\pgfsetbuttcap%
\pgfsetmiterjoin%
\definecolor{currentfill}{rgb}{0.445098,0.715686,0.630392}%
\pgfsetfillcolor{currentfill}%
\pgfsetlinewidth{1.003750pt}%
\definecolor{currentstroke}{rgb}{1.000000,1.000000,1.000000}%
\pgfsetstrokecolor{currentstroke}%
\pgfsetdash{}{0pt}%
\pgfpathmoveto{\pgfqpoint{3.583934in}{-226.701573in}}%
\pgfpathlineto{\pgfqpoint{3.583934in}{2.678563in}}%
\pgfpathlineto{\pgfqpoint{3.445170in}{2.678563in}}%
\pgfpathlineto{\pgfqpoint{3.445170in}{-226.701573in}}%
\pgfusepath{stroke,fill}%
\end{pgfscope}%
\begin{pgfscope}%
\pgfpathrectangle{\pgfqpoint{0.635194in}{0.763790in}}{\pgfqpoint{3.469105in}{2.038210in}}%
\pgfusepath{clip}%
\pgfsetbuttcap%
\pgfsetmiterjoin%
\definecolor{currentfill}{rgb}{0.445098,0.715686,0.630392}%
\pgfsetfillcolor{currentfill}%
\pgfsetlinewidth{1.003750pt}%
\definecolor{currentstroke}{rgb}{1.000000,1.000000,1.000000}%
\pgfsetstrokecolor{currentstroke}%
\pgfsetdash{}{0pt}%
\pgfpathmoveto{\pgfqpoint{3.930844in}{-226.701573in}}%
\pgfpathlineto{\pgfqpoint{3.930844in}{2.709354in}}%
\pgfpathlineto{\pgfqpoint{3.792080in}{2.709354in}}%
\pgfpathlineto{\pgfqpoint{3.792080in}{-226.701573in}}%
\pgfusepath{stroke,fill}%
\end{pgfscope}%
\begin{pgfscope}%
\pgfpathrectangle{\pgfqpoint{0.635194in}{0.763790in}}{\pgfqpoint{3.469105in}{2.038210in}}%
\pgfusepath{clip}%
\pgfsetbuttcap%
\pgfsetmiterjoin%
\definecolor{currentfill}{rgb}{0.912745,0.586275,0.459804}%
\pgfsetfillcolor{currentfill}%
\pgfsetlinewidth{1.003750pt}%
\definecolor{currentstroke}{rgb}{1.000000,1.000000,1.000000}%
\pgfsetstrokecolor{currentstroke}%
\pgfsetdash{}{0pt}%
\pgfpathmoveto{\pgfqpoint{0.947414in}{-226.701573in}}%
\pgfpathlineto{\pgfqpoint{0.947414in}{0.922577in}}%
\pgfpathlineto{\pgfqpoint{0.808650in}{0.922577in}}%
\pgfpathlineto{\pgfqpoint{0.808650in}{-226.701573in}}%
\pgfusepath{stroke,fill}%
\end{pgfscope}%
\begin{pgfscope}%
\pgfpathrectangle{\pgfqpoint{0.635194in}{0.763790in}}{\pgfqpoint{3.469105in}{2.038210in}}%
\pgfusepath{clip}%
\pgfsetbuttcap%
\pgfsetmiterjoin%
\definecolor{currentfill}{rgb}{0.912745,0.586275,0.459804}%
\pgfsetfillcolor{currentfill}%
\pgfsetlinewidth{1.003750pt}%
\definecolor{currentstroke}{rgb}{1.000000,1.000000,1.000000}%
\pgfsetstrokecolor{currentstroke}%
\pgfsetdash{}{0pt}%
\pgfpathmoveto{\pgfqpoint{1.294324in}{-226.701573in}}%
\pgfpathlineto{\pgfqpoint{1.294324in}{1.762529in}}%
\pgfpathlineto{\pgfqpoint{1.155560in}{1.762529in}}%
\pgfpathlineto{\pgfqpoint{1.155560in}{-226.701573in}}%
\pgfusepath{stroke,fill}%
\end{pgfscope}%
\begin{pgfscope}%
\pgfpathrectangle{\pgfqpoint{0.635194in}{0.763790in}}{\pgfqpoint{3.469105in}{2.038210in}}%
\pgfusepath{clip}%
\pgfsetbuttcap%
\pgfsetmiterjoin%
\definecolor{currentfill}{rgb}{0.912745,0.586275,0.459804}%
\pgfsetfillcolor{currentfill}%
\pgfsetlinewidth{1.003750pt}%
\definecolor{currentstroke}{rgb}{1.000000,1.000000,1.000000}%
\pgfsetstrokecolor{currentstroke}%
\pgfsetdash{}{0pt}%
\pgfpathmoveto{\pgfqpoint{1.641235in}{-226.701573in}}%
\pgfpathlineto{\pgfqpoint{1.641235in}{1.871287in}}%
\pgfpathlineto{\pgfqpoint{1.502471in}{1.871287in}}%
\pgfpathlineto{\pgfqpoint{1.502471in}{-226.701573in}}%
\pgfusepath{stroke,fill}%
\end{pgfscope}%
\begin{pgfscope}%
\pgfpathrectangle{\pgfqpoint{0.635194in}{0.763790in}}{\pgfqpoint{3.469105in}{2.038210in}}%
\pgfusepath{clip}%
\pgfsetbuttcap%
\pgfsetmiterjoin%
\definecolor{currentfill}{rgb}{0.912745,0.586275,0.459804}%
\pgfsetfillcolor{currentfill}%
\pgfsetlinewidth{1.003750pt}%
\definecolor{currentstroke}{rgb}{1.000000,1.000000,1.000000}%
\pgfsetstrokecolor{currentstroke}%
\pgfsetdash{}{0pt}%
\pgfpathmoveto{\pgfqpoint{1.988146in}{-226.701573in}}%
\pgfpathlineto{\pgfqpoint{1.988146in}{1.845672in}}%
\pgfpathlineto{\pgfqpoint{1.849381in}{1.845672in}}%
\pgfpathlineto{\pgfqpoint{1.849381in}{-226.701573in}}%
\pgfusepath{stroke,fill}%
\end{pgfscope}%
\begin{pgfscope}%
\pgfpathrectangle{\pgfqpoint{0.635194in}{0.763790in}}{\pgfqpoint{3.469105in}{2.038210in}}%
\pgfusepath{clip}%
\pgfsetbuttcap%
\pgfsetmiterjoin%
\definecolor{currentfill}{rgb}{0.912745,0.586275,0.459804}%
\pgfsetfillcolor{currentfill}%
\pgfsetlinewidth{1.003750pt}%
\definecolor{currentstroke}{rgb}{1.000000,1.000000,1.000000}%
\pgfsetstrokecolor{currentstroke}%
\pgfsetdash{}{0pt}%
\pgfpathmoveto{\pgfqpoint{2.335056in}{-226.701573in}}%
\pgfpathlineto{\pgfqpoint{2.335056in}{1.895697in}}%
\pgfpathlineto{\pgfqpoint{2.196292in}{1.895697in}}%
\pgfpathlineto{\pgfqpoint{2.196292in}{-226.701573in}}%
\pgfusepath{stroke,fill}%
\end{pgfscope}%
\begin{pgfscope}%
\pgfpathrectangle{\pgfqpoint{0.635194in}{0.763790in}}{\pgfqpoint{3.469105in}{2.038210in}}%
\pgfusepath{clip}%
\pgfsetbuttcap%
\pgfsetmiterjoin%
\definecolor{currentfill}{rgb}{0.912745,0.586275,0.459804}%
\pgfsetfillcolor{currentfill}%
\pgfsetlinewidth{1.003750pt}%
\definecolor{currentstroke}{rgb}{1.000000,1.000000,1.000000}%
\pgfsetstrokecolor{currentstroke}%
\pgfsetdash{}{0pt}%
\pgfpathmoveto{\pgfqpoint{2.681967in}{-226.701573in}}%
\pgfpathlineto{\pgfqpoint{2.681967in}{1.969529in}}%
\pgfpathlineto{\pgfqpoint{2.543202in}{1.969529in}}%
\pgfpathlineto{\pgfqpoint{2.543202in}{-226.701573in}}%
\pgfusepath{stroke,fill}%
\end{pgfscope}%
\begin{pgfscope}%
\pgfpathrectangle{\pgfqpoint{0.635194in}{0.763790in}}{\pgfqpoint{3.469105in}{2.038210in}}%
\pgfusepath{clip}%
\pgfsetbuttcap%
\pgfsetmiterjoin%
\definecolor{currentfill}{rgb}{0.912745,0.586275,0.459804}%
\pgfsetfillcolor{currentfill}%
\pgfsetlinewidth{1.003750pt}%
\definecolor{currentstroke}{rgb}{1.000000,1.000000,1.000000}%
\pgfsetstrokecolor{currentstroke}%
\pgfsetdash{}{0pt}%
\pgfpathmoveto{\pgfqpoint{3.028877in}{-226.701573in}}%
\pgfpathlineto{\pgfqpoint{3.028877in}{2.217509in}}%
\pgfpathlineto{\pgfqpoint{2.890113in}{2.217509in}}%
\pgfpathlineto{\pgfqpoint{2.890113in}{-226.701573in}}%
\pgfusepath{stroke,fill}%
\end{pgfscope}%
\begin{pgfscope}%
\pgfpathrectangle{\pgfqpoint{0.635194in}{0.763790in}}{\pgfqpoint{3.469105in}{2.038210in}}%
\pgfusepath{clip}%
\pgfsetbuttcap%
\pgfsetmiterjoin%
\definecolor{currentfill}{rgb}{0.912745,0.586275,0.459804}%
\pgfsetfillcolor{currentfill}%
\pgfsetlinewidth{1.003750pt}%
\definecolor{currentstroke}{rgb}{1.000000,1.000000,1.000000}%
\pgfsetstrokecolor{currentstroke}%
\pgfsetdash{}{0pt}%
\pgfpathmoveto{\pgfqpoint{3.375788in}{-226.701573in}}%
\pgfpathlineto{\pgfqpoint{3.375788in}{2.206641in}}%
\pgfpathlineto{\pgfqpoint{3.237023in}{2.206641in}}%
\pgfpathlineto{\pgfqpoint{3.237023in}{-226.701573in}}%
\pgfusepath{stroke,fill}%
\end{pgfscope}%
\begin{pgfscope}%
\pgfpathrectangle{\pgfqpoint{0.635194in}{0.763790in}}{\pgfqpoint{3.469105in}{2.038210in}}%
\pgfusepath{clip}%
\pgfsetbuttcap%
\pgfsetmiterjoin%
\definecolor{currentfill}{rgb}{0.912745,0.586275,0.459804}%
\pgfsetfillcolor{currentfill}%
\pgfsetlinewidth{1.003750pt}%
\definecolor{currentstroke}{rgb}{1.000000,1.000000,1.000000}%
\pgfsetstrokecolor{currentstroke}%
\pgfsetdash{}{0pt}%
\pgfpathmoveto{\pgfqpoint{3.722698in}{-226.701573in}}%
\pgfpathlineto{\pgfqpoint{3.722698in}{2.619632in}}%
\pgfpathlineto{\pgfqpoint{3.583934in}{2.619632in}}%
\pgfpathlineto{\pgfqpoint{3.583934in}{-226.701573in}}%
\pgfusepath{stroke,fill}%
\end{pgfscope}%
\begin{pgfscope}%
\pgfpathrectangle{\pgfqpoint{0.635194in}{0.763790in}}{\pgfqpoint{3.469105in}{2.038210in}}%
\pgfusepath{clip}%
\pgfsetbuttcap%
\pgfsetmiterjoin%
\definecolor{currentfill}{rgb}{0.912745,0.586275,0.459804}%
\pgfsetfillcolor{currentfill}%
\pgfsetlinewidth{1.003750pt}%
\definecolor{currentstroke}{rgb}{1.000000,1.000000,1.000000}%
\pgfsetstrokecolor{currentstroke}%
\pgfsetdash{}{0pt}%
\pgfpathmoveto{\pgfqpoint{4.069609in}{-226.701573in}}%
\pgfpathlineto{\pgfqpoint{4.069609in}{2.654761in}}%
\pgfpathlineto{\pgfqpoint{3.930844in}{2.654761in}}%
\pgfpathlineto{\pgfqpoint{3.930844in}{-226.701573in}}%
\pgfusepath{stroke,fill}%
\end{pgfscope}%
\begin{pgfscope}%
\pgfpathrectangle{\pgfqpoint{0.635194in}{0.763790in}}{\pgfqpoint{3.469105in}{2.038210in}}%
\pgfusepath{clip}%
\pgfsetroundcap%
\pgfsetroundjoin%
\pgfsetlinewidth{2.710125pt}%
\definecolor{currentstroke}{rgb}{0.260000,0.260000,0.260000}%
\pgfsetstrokecolor{currentstroke}%
\pgfsetdash{}{0pt}%
\pgfusepath{stroke}%
\end{pgfscope}%
\begin{pgfscope}%
\pgfpathrectangle{\pgfqpoint{0.635194in}{0.763790in}}{\pgfqpoint{3.469105in}{2.038210in}}%
\pgfusepath{clip}%
\pgfsetroundcap%
\pgfsetroundjoin%
\pgfsetlinewidth{2.710125pt}%
\definecolor{currentstroke}{rgb}{0.260000,0.260000,0.260000}%
\pgfsetstrokecolor{currentstroke}%
\pgfsetdash{}{0pt}%
\pgfusepath{stroke}%
\end{pgfscope}%
\begin{pgfscope}%
\pgfpathrectangle{\pgfqpoint{0.635194in}{0.763790in}}{\pgfqpoint{3.469105in}{2.038210in}}%
\pgfusepath{clip}%
\pgfsetroundcap%
\pgfsetroundjoin%
\pgfsetlinewidth{2.710125pt}%
\definecolor{currentstroke}{rgb}{0.260000,0.260000,0.260000}%
\pgfsetstrokecolor{currentstroke}%
\pgfsetdash{}{0pt}%
\pgfusepath{stroke}%
\end{pgfscope}%
\begin{pgfscope}%
\pgfpathrectangle{\pgfqpoint{0.635194in}{0.763790in}}{\pgfqpoint{3.469105in}{2.038210in}}%
\pgfusepath{clip}%
\pgfsetroundcap%
\pgfsetroundjoin%
\pgfsetlinewidth{2.710125pt}%
\definecolor{currentstroke}{rgb}{0.260000,0.260000,0.260000}%
\pgfsetstrokecolor{currentstroke}%
\pgfsetdash{}{0pt}%
\pgfusepath{stroke}%
\end{pgfscope}%
\begin{pgfscope}%
\pgfpathrectangle{\pgfqpoint{0.635194in}{0.763790in}}{\pgfqpoint{3.469105in}{2.038210in}}%
\pgfusepath{clip}%
\pgfsetroundcap%
\pgfsetroundjoin%
\pgfsetlinewidth{2.710125pt}%
\definecolor{currentstroke}{rgb}{0.260000,0.260000,0.260000}%
\pgfsetstrokecolor{currentstroke}%
\pgfsetdash{}{0pt}%
\pgfusepath{stroke}%
\end{pgfscope}%
\begin{pgfscope}%
\pgfpathrectangle{\pgfqpoint{0.635194in}{0.763790in}}{\pgfqpoint{3.469105in}{2.038210in}}%
\pgfusepath{clip}%
\pgfsetroundcap%
\pgfsetroundjoin%
\pgfsetlinewidth{2.710125pt}%
\definecolor{currentstroke}{rgb}{0.260000,0.260000,0.260000}%
\pgfsetstrokecolor{currentstroke}%
\pgfsetdash{}{0pt}%
\pgfusepath{stroke}%
\end{pgfscope}%
\begin{pgfscope}%
\pgfpathrectangle{\pgfqpoint{0.635194in}{0.763790in}}{\pgfqpoint{3.469105in}{2.038210in}}%
\pgfusepath{clip}%
\pgfsetroundcap%
\pgfsetroundjoin%
\pgfsetlinewidth{2.710125pt}%
\definecolor{currentstroke}{rgb}{0.260000,0.260000,0.260000}%
\pgfsetstrokecolor{currentstroke}%
\pgfsetdash{}{0pt}%
\pgfusepath{stroke}%
\end{pgfscope}%
\begin{pgfscope}%
\pgfpathrectangle{\pgfqpoint{0.635194in}{0.763790in}}{\pgfqpoint{3.469105in}{2.038210in}}%
\pgfusepath{clip}%
\pgfsetroundcap%
\pgfsetroundjoin%
\pgfsetlinewidth{2.710125pt}%
\definecolor{currentstroke}{rgb}{0.260000,0.260000,0.260000}%
\pgfsetstrokecolor{currentstroke}%
\pgfsetdash{}{0pt}%
\pgfusepath{stroke}%
\end{pgfscope}%
\begin{pgfscope}%
\pgfpathrectangle{\pgfqpoint{0.635194in}{0.763790in}}{\pgfqpoint{3.469105in}{2.038210in}}%
\pgfusepath{clip}%
\pgfsetroundcap%
\pgfsetroundjoin%
\pgfsetlinewidth{2.710125pt}%
\definecolor{currentstroke}{rgb}{0.260000,0.260000,0.260000}%
\pgfsetstrokecolor{currentstroke}%
\pgfsetdash{}{0pt}%
\pgfusepath{stroke}%
\end{pgfscope}%
\begin{pgfscope}%
\pgfpathrectangle{\pgfqpoint{0.635194in}{0.763790in}}{\pgfqpoint{3.469105in}{2.038210in}}%
\pgfusepath{clip}%
\pgfsetroundcap%
\pgfsetroundjoin%
\pgfsetlinewidth{2.710125pt}%
\definecolor{currentstroke}{rgb}{0.260000,0.260000,0.260000}%
\pgfsetstrokecolor{currentstroke}%
\pgfsetdash{}{0pt}%
\pgfusepath{stroke}%
\end{pgfscope}%
\begin{pgfscope}%
\pgfpathrectangle{\pgfqpoint{0.635194in}{0.763790in}}{\pgfqpoint{3.469105in}{2.038210in}}%
\pgfusepath{clip}%
\pgfsetroundcap%
\pgfsetroundjoin%
\pgfsetlinewidth{2.710125pt}%
\definecolor{currentstroke}{rgb}{0.260000,0.260000,0.260000}%
\pgfsetstrokecolor{currentstroke}%
\pgfsetdash{}{0pt}%
\pgfusepath{stroke}%
\end{pgfscope}%
\begin{pgfscope}%
\pgfpathrectangle{\pgfqpoint{0.635194in}{0.763790in}}{\pgfqpoint{3.469105in}{2.038210in}}%
\pgfusepath{clip}%
\pgfsetroundcap%
\pgfsetroundjoin%
\pgfsetlinewidth{2.710125pt}%
\definecolor{currentstroke}{rgb}{0.260000,0.260000,0.260000}%
\pgfsetstrokecolor{currentstroke}%
\pgfsetdash{}{0pt}%
\pgfusepath{stroke}%
\end{pgfscope}%
\begin{pgfscope}%
\pgfpathrectangle{\pgfqpoint{0.635194in}{0.763790in}}{\pgfqpoint{3.469105in}{2.038210in}}%
\pgfusepath{clip}%
\pgfsetroundcap%
\pgfsetroundjoin%
\pgfsetlinewidth{2.710125pt}%
\definecolor{currentstroke}{rgb}{0.260000,0.260000,0.260000}%
\pgfsetstrokecolor{currentstroke}%
\pgfsetdash{}{0pt}%
\pgfusepath{stroke}%
\end{pgfscope}%
\begin{pgfscope}%
\pgfpathrectangle{\pgfqpoint{0.635194in}{0.763790in}}{\pgfqpoint{3.469105in}{2.038210in}}%
\pgfusepath{clip}%
\pgfsetroundcap%
\pgfsetroundjoin%
\pgfsetlinewidth{2.710125pt}%
\definecolor{currentstroke}{rgb}{0.260000,0.260000,0.260000}%
\pgfsetstrokecolor{currentstroke}%
\pgfsetdash{}{0pt}%
\pgfusepath{stroke}%
\end{pgfscope}%
\begin{pgfscope}%
\pgfpathrectangle{\pgfqpoint{0.635194in}{0.763790in}}{\pgfqpoint{3.469105in}{2.038210in}}%
\pgfusepath{clip}%
\pgfsetroundcap%
\pgfsetroundjoin%
\pgfsetlinewidth{2.710125pt}%
\definecolor{currentstroke}{rgb}{0.260000,0.260000,0.260000}%
\pgfsetstrokecolor{currentstroke}%
\pgfsetdash{}{0pt}%
\pgfusepath{stroke}%
\end{pgfscope}%
\begin{pgfscope}%
\pgfpathrectangle{\pgfqpoint{0.635194in}{0.763790in}}{\pgfqpoint{3.469105in}{2.038210in}}%
\pgfusepath{clip}%
\pgfsetroundcap%
\pgfsetroundjoin%
\pgfsetlinewidth{2.710125pt}%
\definecolor{currentstroke}{rgb}{0.260000,0.260000,0.260000}%
\pgfsetstrokecolor{currentstroke}%
\pgfsetdash{}{0pt}%
\pgfusepath{stroke}%
\end{pgfscope}%
\begin{pgfscope}%
\pgfpathrectangle{\pgfqpoint{0.635194in}{0.763790in}}{\pgfqpoint{3.469105in}{2.038210in}}%
\pgfusepath{clip}%
\pgfsetroundcap%
\pgfsetroundjoin%
\pgfsetlinewidth{2.710125pt}%
\definecolor{currentstroke}{rgb}{0.260000,0.260000,0.260000}%
\pgfsetstrokecolor{currentstroke}%
\pgfsetdash{}{0pt}%
\pgfusepath{stroke}%
\end{pgfscope}%
\begin{pgfscope}%
\pgfpathrectangle{\pgfqpoint{0.635194in}{0.763790in}}{\pgfqpoint{3.469105in}{2.038210in}}%
\pgfusepath{clip}%
\pgfsetroundcap%
\pgfsetroundjoin%
\pgfsetlinewidth{2.710125pt}%
\definecolor{currentstroke}{rgb}{0.260000,0.260000,0.260000}%
\pgfsetstrokecolor{currentstroke}%
\pgfsetdash{}{0pt}%
\pgfusepath{stroke}%
\end{pgfscope}%
\begin{pgfscope}%
\pgfpathrectangle{\pgfqpoint{0.635194in}{0.763790in}}{\pgfqpoint{3.469105in}{2.038210in}}%
\pgfusepath{clip}%
\pgfsetroundcap%
\pgfsetroundjoin%
\pgfsetlinewidth{2.710125pt}%
\definecolor{currentstroke}{rgb}{0.260000,0.260000,0.260000}%
\pgfsetstrokecolor{currentstroke}%
\pgfsetdash{}{0pt}%
\pgfusepath{stroke}%
\end{pgfscope}%
\begin{pgfscope}%
\pgfpathrectangle{\pgfqpoint{0.635194in}{0.763790in}}{\pgfqpoint{3.469105in}{2.038210in}}%
\pgfusepath{clip}%
\pgfsetroundcap%
\pgfsetroundjoin%
\pgfsetlinewidth{2.710125pt}%
\definecolor{currentstroke}{rgb}{0.260000,0.260000,0.260000}%
\pgfsetstrokecolor{currentstroke}%
\pgfsetdash{}{0pt}%
\pgfusepath{stroke}%
\end{pgfscope}%
\begin{pgfscope}%
\pgfsetrectcap%
\pgfsetmiterjoin%
\pgfsetlinewidth{1.254687pt}%
\definecolor{currentstroke}{rgb}{0.800000,0.800000,0.800000}%
\pgfsetstrokecolor{currentstroke}%
\pgfsetdash{}{0pt}%
\pgfpathmoveto{\pgfqpoint{0.635194in}{0.763790in}}%
\pgfpathlineto{\pgfqpoint{0.635194in}{2.802000in}}%
\pgfusepath{stroke}%
\end{pgfscope}%
\begin{pgfscope}%
\pgfsetrectcap%
\pgfsetmiterjoin%
\pgfsetlinewidth{1.254687pt}%
\definecolor{currentstroke}{rgb}{0.800000,0.800000,0.800000}%
\pgfsetstrokecolor{currentstroke}%
\pgfsetdash{}{0pt}%
\pgfpathmoveto{\pgfqpoint{0.635194in}{0.763790in}}%
\pgfpathlineto{\pgfqpoint{4.104300in}{0.763790in}}%
\pgfusepath{stroke}%
\end{pgfscope}%
\begin{pgfscope}%
\pgfsetbuttcap%
\pgfsetmiterjoin%
\definecolor{currentfill}{rgb}{0.445098,0.715686,0.630392}%
\pgfsetfillcolor{currentfill}%
\pgfsetlinewidth{1.003750pt}%
\definecolor{currentstroke}{rgb}{1.000000,1.000000,1.000000}%
\pgfsetstrokecolor{currentstroke}%
\pgfsetdash{}{0pt}%
\pgfpathmoveto{\pgfqpoint{1.705928in}{2.925111in}}%
\pgfpathlineto{\pgfqpoint{1.928150in}{2.925111in}}%
\pgfpathlineto{\pgfqpoint{1.928150in}{3.002889in}}%
\pgfpathlineto{\pgfqpoint{1.705928in}{3.002889in}}%
\pgfpathlineto{\pgfqpoint{1.705928in}{2.925111in}}%
\pgfpathclose%
\pgfusepath{stroke,fill}%
\end{pgfscope}%
\begin{pgfscope}%
\definecolor{textcolor}{rgb}{0.150000,0.150000,0.150000}%
\pgfsetstrokecolor{textcolor}%
\pgfsetfillcolor{textcolor}%
\pgftext[x=2.017039in,y=2.925111in,left,base]{\color{textcolor}\sffamily\fontsize{8.000000}{9.600000}\selectfont \(\displaystyle TP_U\)}%
\end{pgfscope}%
\begin{pgfscope}%
\pgfsetbuttcap%
\pgfsetmiterjoin%
\definecolor{currentfill}{rgb}{0.912745,0.586275,0.459804}%
\pgfsetfillcolor{currentfill}%
\pgfsetlinewidth{1.003750pt}%
\definecolor{currentstroke}{rgb}{1.000000,1.000000,1.000000}%
\pgfsetstrokecolor{currentstroke}%
\pgfsetdash{}{0pt}%
\pgfpathmoveto{\pgfqpoint{2.485011in}{2.925111in}}%
\pgfpathlineto{\pgfqpoint{2.707234in}{2.925111in}}%
\pgfpathlineto{\pgfqpoint{2.707234in}{3.002889in}}%
\pgfpathlineto{\pgfqpoint{2.485011in}{3.002889in}}%
\pgfpathlineto{\pgfqpoint{2.485011in}{2.925111in}}%
\pgfpathclose%
\pgfusepath{stroke,fill}%
\end{pgfscope}%
\begin{pgfscope}%
\definecolor{textcolor}{rgb}{0.150000,0.150000,0.150000}%
\pgfsetstrokecolor{textcolor}%
\pgfsetfillcolor{textcolor}%
\pgftext[x=2.796123in,y=2.925111in,left,base]{\color{textcolor}\sffamily\fontsize{8.000000}{9.600000}\selectfont \(\displaystyle TP_L\)}%
\end{pgfscope}%
\end{pgfpicture}%
\makeatother%
\endgroup%